\documentclass[mathleft]{an}
\usepackage{graphicx}
\usepackage{times}
\usepackage{aas_macros}
\usepackage{amsmath}
\usepackage{natbib}
\bibpunct{(}{)}{;}{a}{}{,}

\begin{document}

\Pagespan{1}{}
\Yearpublication{2013}
\Yearsubmission{2013}
\Month{1}
\Volume{XXX}
\Issue{XXX}

\title{Modelling Giant Radio Halos}

\author{J.M.F. Donnert\inst{1}\fnmsep\thanks{\email{donnert@ira.inaf.it}}}

\titlerunning{Giant Radio Haloes}
\authorrunning{J.M.F. Donnert}
\institute{Istituto di Radioastronomia, via P. Gobetti 101, 40129 Bologna, ER, Italy}

\received{xx xxx 2013}
\accepted{xx xxx 2013}
\publonline{xxxxx}

\keywords{galaxies: clusters, radio continuum: general, radiation mechanism: non-thermal}

\abstract{We review models for giant radio halos in clusters of galaxies, with a focus on numerical and theoretical work. After summarising the most important observations of these objects, we present an introduction to the theoretical aspects of hadronic models. We compare these models with observations using simulations and find severe problems for hadronic models. We give a short introduction to reacceleration models and show results from the first simulation of CRe reacceleration in cluster mergers. We find that in-line with previous theoretical work, reacceleration models are able to elegantly explain main observables of giant radio halos.}

\maketitle

\section{Introduction}\label{intro}

Galaxy clusters form the knots of the cosmic web through merging of smaller structures seeded as density fluctuations in the early Universe. Despite the name, only a few percent of a clusters mass is accounted for by actual galaxies - most of it is in collisionless dark matter, which dominates the gravitational potential. The merging process deposits baryons in this potential which form the intra-cluster-medium (ICM). During large mergers this infall can dissipate up to $10^{63}\,\mathrm{erg}$ of kinetic energy in the ICM into turbulence, shocks, relativistic particles and eventually heat \citep[e.g.][]{2006MNRAS.366.1437S}. \par
Due to its high temperatures of more than $10^8 \, \mathrm{K}$ the thermal bremsstrahlung of the ICM plasma is prominently observed in the X-rays \citep{1971Natur.231..107M,1971Natur.231..437C}. However the ICM was first discovered through radio observations of the Coma cluster \citep{1970MNRAS.151....1W}. Here the interaction of relativistic electrons with the magnetic field of the plasma produces synchrotron radiation - a process well known from lobes of radio galaxies \citep{1968MNRAS.138....1R}. This proves the presence of non-thermal components in the ICM: cosmic-ray electrons (CRe) and magnetic fields. Later turbulence was independently estimated in Coma through observed temperature fluctuations \citep{2004A&A...426..387S}.\par
\begin{figure}
	\centering
	\includegraphics[width=0.4\textwidth]{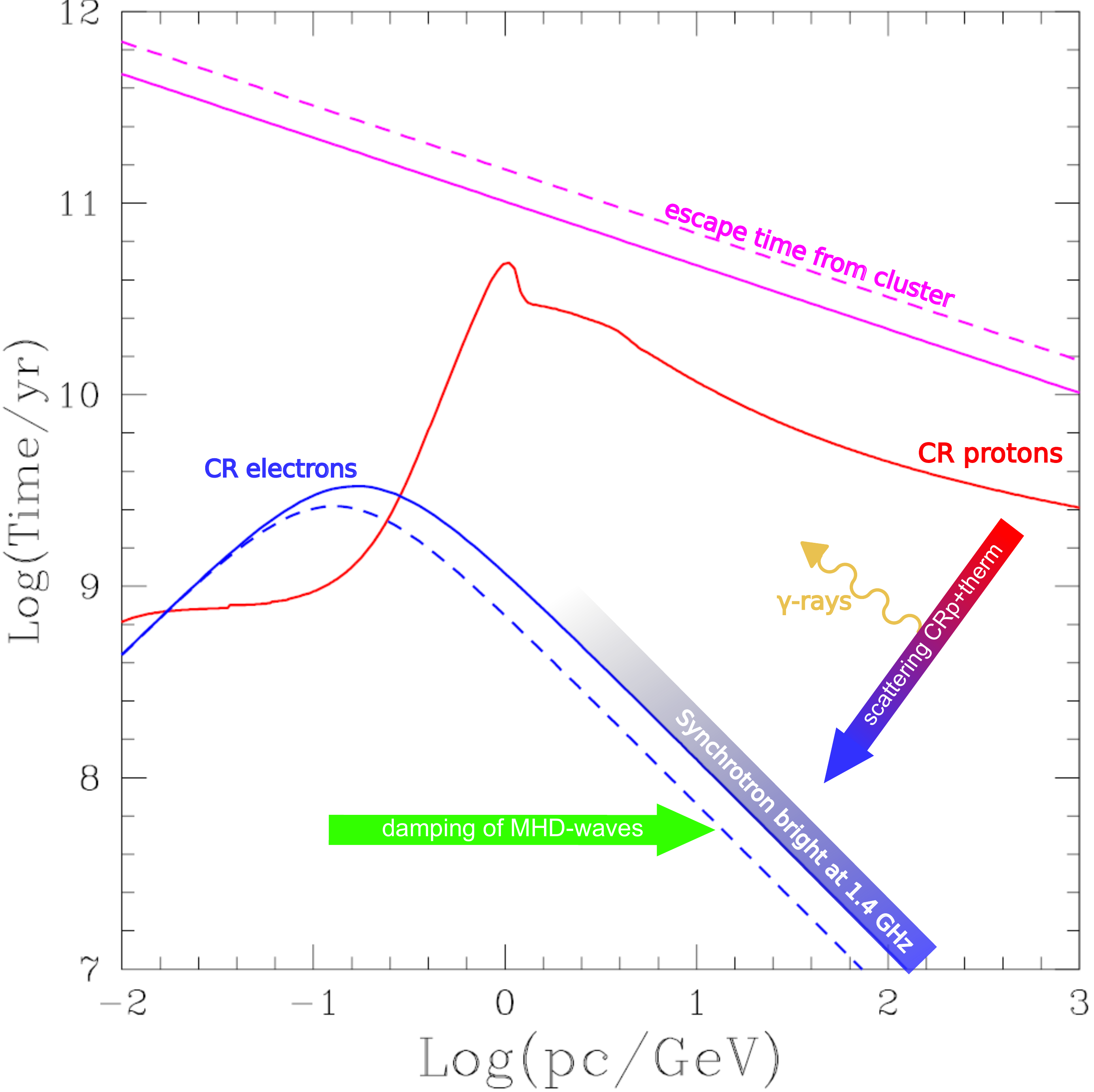}
	\caption{Lifetime of CR electrons (blue) and CR protons (red) in the ICM as well as their escape time (pink). Plotted for magnetic field strength of $1\,\mu\mathrm{G}$ (full) and $0.1\,\mu\mathrm{G}$ (dashed). Synchrotron bright energies are marked blue. Figure adapted and modified from \citet{2007IJMPA..22..681B}. }
	\label{img.coolingtime}
\end{figure}
\begin{figure*}
	\centering
	\includegraphics[width=0.9\textwidth]{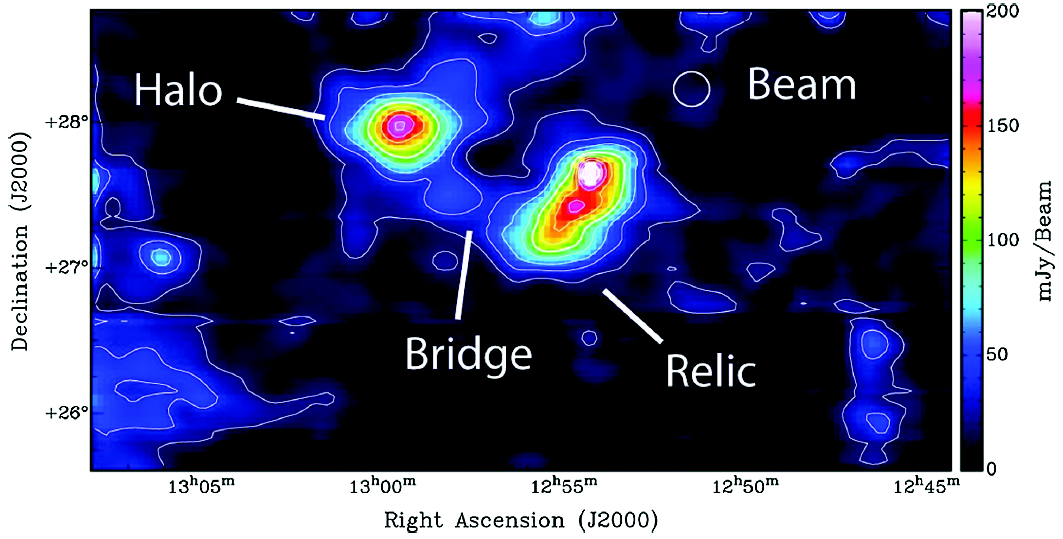}
	\caption{Complex non-thermal emission observed with GBT at 1.41 Ghz in Coma by \citet{2011MNRAS.412....2B}. The beam size is roughly 360 kpc. Contours start at  20 mJy/beam and increase by 20 mJy/beam.}
	\label{img.coma}
\end{figure*}

Today non-thermal phenomena in galaxy clusters are classified as  radio halos, radio relics and mini halos \citep{1996IAUS..175..333F}. \emph{Radio halos} are centered diffuse unpolarised objects found exclusively in clusters with disturbed X-ray morphology. \emph{Radio relics} are thin, elongated, polarised radio structures offset from the cluster center.  Both previous classes extend for one to two Mpc in size, while \emph{radio mini halos} are usually limited to a few hundred kpc in the center of relaxed clusters. In this paper we focus on radio halos.\par
Independent measurements of the ICM magnetic field through Faraday rotation of intrinsic polarised radio sources \citep[e.g.][]{2010A&A...513A..30B} reveal field strengths of up to a few $\mu\mathrm{G}$ in the center of clusters, declining outwards. The field is observed to be highly tangled \citep{2011A&A...529A..13K}, which is expected for a turbulent high beta plasma.\par
This is consistent with simulations, which predict efficient amplification by MHD turbulence as substructures merge with a parent cluster \citep{2001A&A...378..777D,2005MNRAS.364..753D,2009MNRAS.398.1678D}. However \citet{2010NatPh...6..520P} claim a radial magnetic distribution in the Virgo cluster. The origins of cluster magnetic fields are still debated \citep{2002RvMP...74..775W}.\par
Considering radio halos the observed field strengths imply a population of synchrotron bright CRe with energies of $>100\, \mathrm{GeV}$ in the whole cluster volume. It is well established that CRe are injected in shocks with Mach numbers larger than 5 through observations of supernova remnands \citep[e.g.][]{2011ApJ...728L..28E}. And on smaller scales AGN and SN shocks\footnote{through galactic outflows} indeed contribute to the CRe content of a cluster. In contrast, accretion shocks during major mergers of clusters have low Mach numbers of  $\approx 3$. The exact mechanism that might lead to an injection in these shocks is still debated \citep{2008ApJ...682L...5S,2010MNRAS.402.2807B,2012ApJ...744...67G}. These primary CRe's are believed to cause radio relics, which are associated with merger shocks in observations.  However due to their geometry none of these mechanisms are volume filling. Therefore CRe  would have to diffuse hundreds of kpc to form a radio halo.\par
Due to the low ambient thermal density of $n_\mathrm{th}<10^{-2}\,\mathrm{cm}^{-3}$, CRe have a classical mean-free-path of kpc in the ICM \citep{1956pfig.book.....S}. However particle motion faster than the Alv\'en speed in a plasma is known to cause plasma waves through the streaming instability \citep{1974ARA&A..12...71W}. These fluctuations in the plasma e-m-field themselves act as scattering agents in the medium and isotropise CR motions. Therefore CRe diffuse in a random walk through the cluster atmosphere, limiting the effective diffusion speed to the Alv\'en speed of $\approx 100 \,\mathrm{km}/\mathrm{s}$. \par
Relativistic electrons are subject to a number of energy losses in the ICM  as well \citep{1994hea2.book.....L}. The most important being inverse Compton scattering (IC), synchrotron radiation, bremsstrahlung and Coulomb scattering. Thus one can define an approximate life-time of these particles at an momentum $p$ as 
\begin{align}
    t_\mathrm{life} &= \frac{p}{\left|\frac{\mathrm{d}p}{\mathrm{d}t}\right|_\mathrm{loss}}. \label{eqn.cooltime}
\end{align}
In figure \ref{img.coolingtime} we show this life-time  for CRe and relativistic protons (CRp) as well as the escape time from the central ICM \citep[modified from][]{2007astro.ph..1545B}. \par

It has been realised early on that these losses limit the diffusion length of radio bright CRe in the ICM to a few 10 kpc at best. \emph{It is unclear how a cluster-wide population of CRe can be maintained to produce a giant radio halo} \citep{1977ApJ...212....1J}. This is known as the life-time problem. \par
The existence of giant radio halos underlines the incompleteness of the physical picture described above. Indeed \citet{1980ApJ...239L..93D} pointed out that  relativistic protons have sufficiently long life-times to fill the cluster volume (see figure \ref{img.coolingtime}). CRp are likely to be injected alongside CRe into the ICM and can themselves inject synchrotron bright CRe in-situ through a hadronic cascade. Models relying on this mechanism are called \emph{hadronic or secondary models} (see section \ref{sect.hadr}).\par
In addition turbulent reacceleration might be able to explain giant radio halos \citep{1977ApJ...212....1J,2001ApJ...557..560P,2001MNRAS.320..365B}. At low energies the synchrotron dark population of CRe with life-times of $>10^9\,\mathrm{yrs}$ can undergo stochastic reacceleration through damping of merger injected turbulence. This Fermi II like mechanism \citep{1949PhRv...75.1169F} lets CRe diffuse to higher momenta creating volume filling radio bright CRe. It may act on CRp as well \citep{2005MNRAS.363.1173B,2011MNRAS.410..127B}. Effectively this mechanism can boost the radio brightness of a CRe population without altering the total CRe density by producing deviations from power-law spectra. This ansatz is realised in \emph{reacceleration models} (see section \ref{sect.reacc}). \par
Today the clear distinction between the two model classes can be considered a historic artefact as both processes certainly take place in the ICM to some degree. However, CRp have not been observed directly in the ICM so far and the efficiency of CRp injection in large scale ICM shocks is unclear. Hybrid particle in cell simulations suggest efficient CRp or CRe injection to be depending on the alignment of the magnetic field to the plane of the shock \citep{2012ApJ...744...67G}. However, the low densities and mach numbers of large shocks in the ICM make these simulations just barely feasable. To this end the efficiency of CRp injection processes has not been firmly established.  \par
In general the plasma physics in the exotic regime of the ICM presents a serious challenge to theorists and simulators. It should be considered that with mean free paths of $>1 \,\mathrm{kpc}$ and collision times $> 10^6 \,\mathrm{yrs}$ in the ICM \citep[e.g.][]{1956pfig.book.....S,1999astro.ph.11439S}  particle-particle interactions are close to time and length scales relevant to radio halos. Therefore collisionality on kpc scales can not be mediated by particle-particle interactions\footnote{It comes at no surprise that hadronic models run into problems with normalisation as we will see later.}. But observations of cold fronts and shocks confirm that the ICM is indeed a collisional plasma on smaller scales \citep{2007PhR...443....1M}. Collisionality can be recovered by plasma waves which results in very high Reynolds numbers of the ICM of $>1000$ or even $>10^5$ \citep{2011MNRAS.412..817B}.  Interestingly field-particle interactions happen on much smaller length and time scales in the ICM (debye length $\approx 2 \times 10^6 \,\mathrm{cm}$ \citep{2002cra..book.....S}). \par
To date the Coma cluster remains the prototypical cluster with non-thermal radio emission (figure \ref{img.coolingtime}, left). It has been studied extensively in the past and detailed studies on other clusters have been somewhat limited \citep{2011MmSAI..82..499V}. However the situation will fundamentaly change with the arrival of new instruments like EMU+WODAN, LOFAR and later the SKA \citep{2010A&A...509A..68C,2012arXiv1210.1020C}. At low frequencies of less than 100 MHz, radio halos are expected to be bright and ubiquitous, due to their steep emission spectrum. As increasingly more LOFAR stations see first light, studies of theoretical models for radio halos are particulary timely.    \par
In the past decade theoretical research on halos has been devided along the lines of the two classical models. Numerical approaches used CRp dynamics and hadronic injection, sometimes with shock injection \citep{2001ApJ...562..233M,2000A&A...362..151D,2007A&A...473...41E,2008MNRAS.385.1211P,2010MNRAS.409..449P,2012MNRAS.421.3375V,2004MNRAS.347..389H,2008MNRAS.391.1511H,2012MNRAS.425L..76B,2010ApJ...722..737K}. Independently there were (semi-)analytic approaches on reacceleration models, sometimes including CRp \citep{2001ApJ...557..560P,2001MNRAS.320..365B,2005MNRAS.357.1313C,2005MNRAS.363.1173B,2007MNRAS.378..245B,2011MNRAS.410..127B}. Turbulence and magnetic fields have been studied extensively in the cluster context as well \citep{2006MNRAS.366.1437S,1999A&A...348..351D, 2008arXiv0808.0609V,2009A&A...504...33V,2011A&A...529A..17V,2008MNRAS.388.1089I,2009ApJ...707...40M,2011MNRAS.tmp..483I,2008Sci...320..909R}. However hadronic models have only recently been compared in detail to observations, while more complete approaches had not been considered in simulations yet \citep{2010MNRAS.407.1565D,2010MNRAS.401...47D}. \par
This report is organised as follows: We start with a general overview of the important observations of radio halos. We then show results from the comparison of hadronic models with these observations. We continue with an introduction to reacceleration models and conclude with the first simulations of CRe reacceleration in direct cluster collisions.   

\section{Observed Properties of Giant Radio Halos}\label{sect.obs}

\begin{figure*}
	\centering
	\includegraphics[height=0.2\textheight]{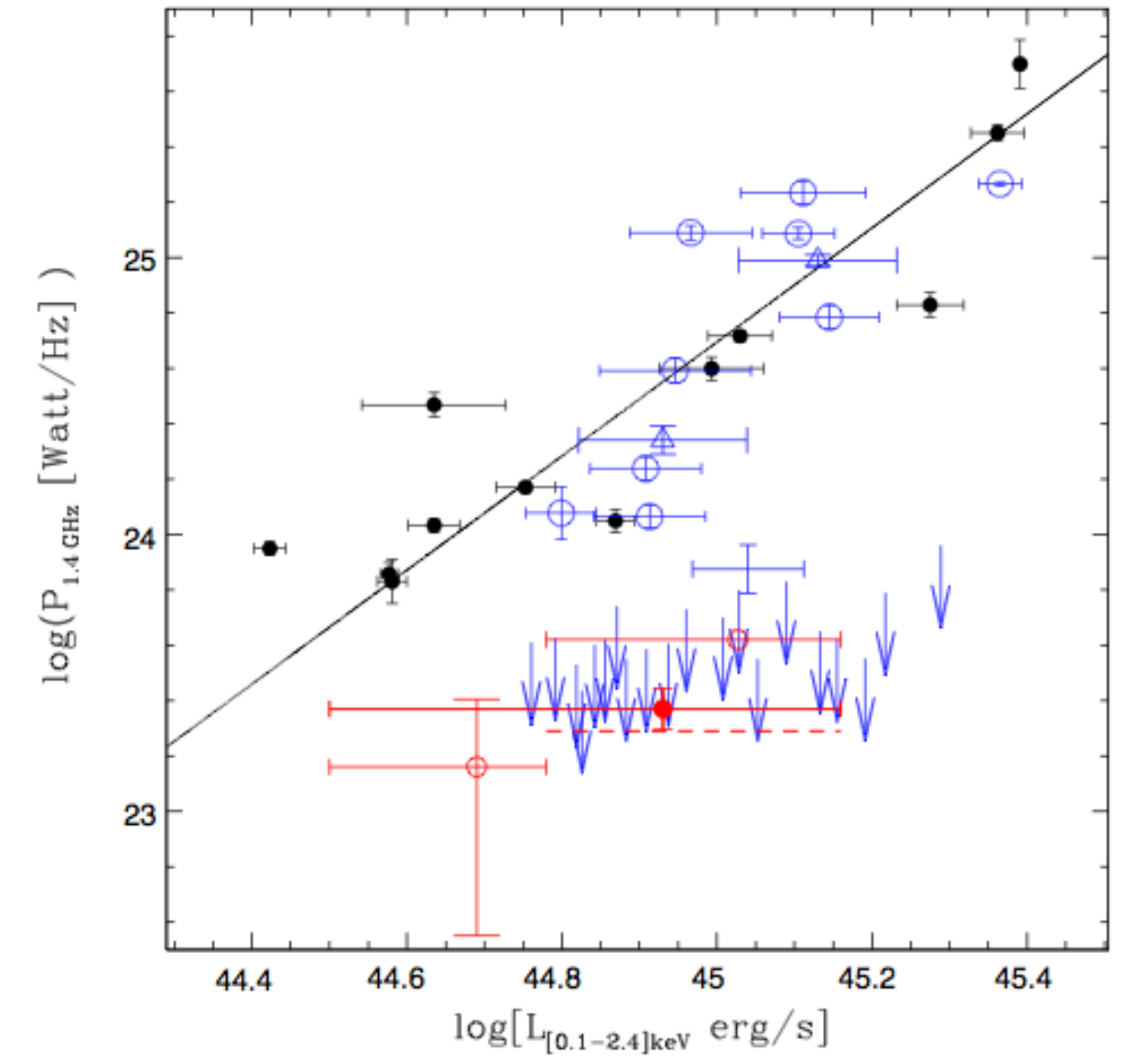}
	\includegraphics[height=0.2\textheight]{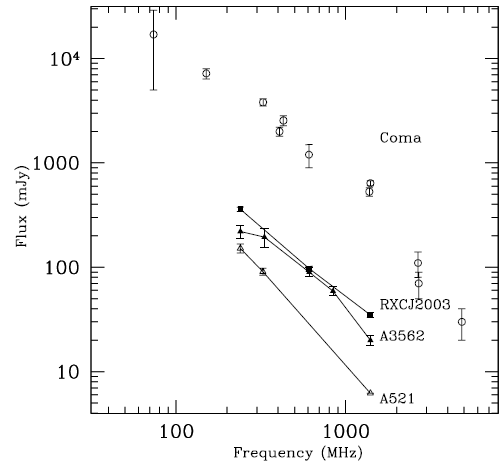}
	\includegraphics[height=0.2\textheight]{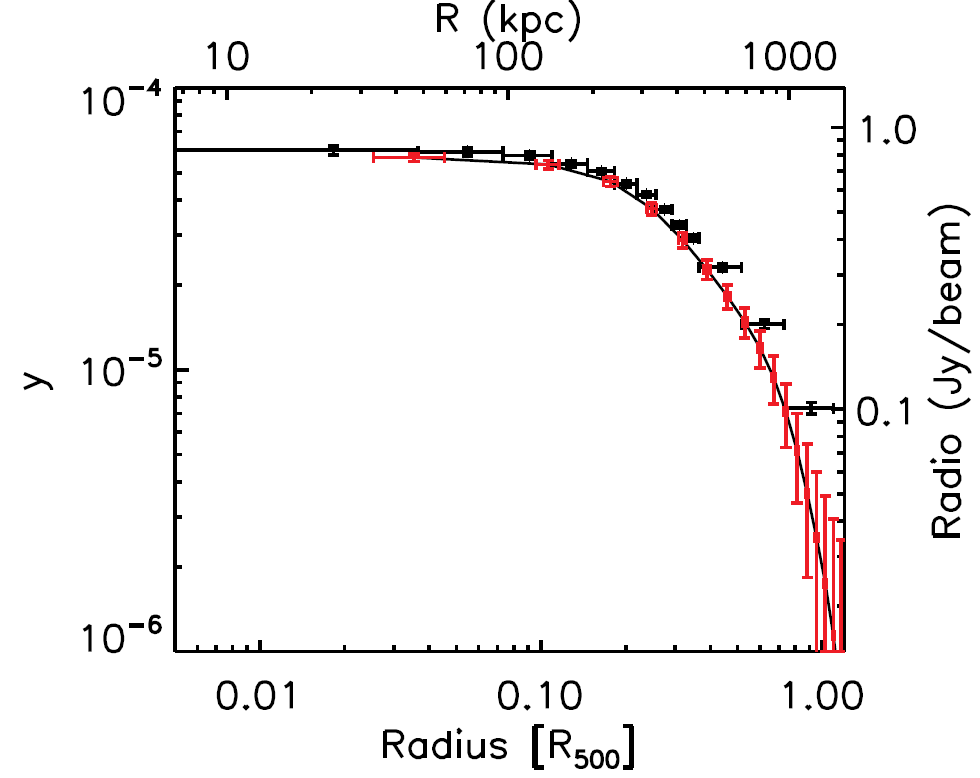}
	\caption{Left: Radio brightness at 1.4 GHz over bolometric X-ray luminosity for a sample of galaxy cluster. Upper limits of non-detections are marked as blue arrows, tentative detection from stacking as red limits \citep[see][ and references therein]{2011ApJ...740L..28B}. Middle: Radio spectra of 4 clusters \citep{2011MmSAI..82..499V}. Right: Radial profile of Compton-y parameter (pressure along the LoS) and radio synchrotron emission in the Coma cluster out to $R_{500}$ \citep{2012arXiv1208.3611P}. }\label{img.observations}
\end{figure*}
Today roughly 50 radio halos are known, mostly at frequencies of 600 MHz and up \citep{2009A&A...507.1257G,2012A&ARv..20...54F}. These halos are all hosted in massive systems with disturbed X-ray morphology \citep{2010ApJ...721L..82C}. These systems form a correlation  in X-ray luminosity and radio brightness with a slope of $\approx 1.8$ (figure \ref{img.observations} left panel). However in a complete X-ray sample these make only 30\% of all systems \citep{2007A&A...463..937V}. Therefore for most clusters only upper limits in the radio are known (arrows in figure \ref{img.observations}, left). Recently a first detection of these clusters has been claimed through a stacking analysis \citep{2011ApJ...740L..28B}.  \par
In the $\gamma$-ray regime, clusters remain not observed \citep{2008AIPC.1085..569P,2009A&A...502..437A,2010ApJ...717L..71A,2012ApJ...757..123A}. \par
The Coma cluster has beed studied extensively in the past. Three observations at 1.4 GHz are commonly used to constrain radio synchrotron profiles of the cluster. It was debated \citep{2004MNRAS.352...76P} wether to use the steeper single-dish profiles from \citet{1997A&A...321...55D} or the flatter interferometer profiles from \citet{2001A&A...376..803G}. However new single-dish GBT observations by \citet{2011MNRAS.412....2B} confirm the flat profile found previously by Govoni and present the deepest profile currently known. \par
For the Coma cluster \citep{2003A&A...397...53T} and a small number of other systems the spectrum of the diffuse emission is observed to be a power-law with a spectral index of 1.2 to 2.5. In figure \ref{img.observations}, middle we show a few known examples from \citet{2011MmSAI..82..499V}. Halos with a synchrotron spectral index $\alpha_\nu > 1.5$ (e.g. A521) are commonly named ultra-steep spectrum haloes. The Spectrum of the Coma cluster shows a \emph{break/steepening} of the emission, which has important theoretical implications (see section \ref{sect.theory}). However differences in instrumentation and analysis might contribute to the total shape of the spectrum. Individual points were estimated with single-dish instruments as well as interferometers. While interferometers might lose diffuse flux because of limited UV-coverage, point source subtraction is complicated for single-dish data. Therefore the spectrum should be taken with a grain of salt. \par
The PLANCK satellite observed the Sunyaev-Zeldovich effect in a numebr clusters with unprecedented sensitivity and resolution. A correlation between the cluster Compton-y and radio luminosity was found from PLANCK data by \citet{2012MNRAS.421L.112B}, however without a clear indication for a bimodal distribution. This suggests that selection effects might contribute to the bimodality found in the X-ray-radio correlation. \par
In Coma, \citet{2012arXiv1208.3611P} found that  pressure and radio emission  are tightly correlated up to $r_{500} \approx 1\,\mathrm{Mpc}$ (figure \ref{img.observations}, right). Previous studies had shown that density and radio luminosity are correlated as well, however the scaling differs between clusters \citep{2001A&A...376..803G}.\par
Recent X-ray observations find a non-thermal pressure of $\le 10\%$ in Coma \citep{2012MNRAS.421.1123C}.

\subsection{Direct Constrains on Models}\label{sect.theory}
The current observational status implies a number of constrains on the available models for a cluster-wide radio bright population of CRe : 
\begin{enumerate}
	\item The predicted radio emission has to be consistent with the observed profile \citep[e.g.][]{2011MNRAS.412....2B}. 
    \item The non-thermal pressure by CRp, magnetic fields and turbulence in clusters may not exceed the non-thermal pressure constraint from the X-rays: $10\%$.
	\item The observation of ultra steep spectrum halos like A512  indicates very steep CRe spectra of $>2$ \citep{2008Natur.455..944B}. This implies either "old" reaccelerated CRe or a CRp population with very steep spectral index in these clusters.  
    \item The non-detection of $\gamma$-rays constrains the amount of CRp in clusters.
	\item The large number of non-detections in the radio regime suggests that radio halos are \emph{transient} phenomena. Statistics roughly indicates that given the observed population a cluster may not stay on the correlation (i.e. be radio loud at 1.4 Ghz) for more than half a Gyr, with a transition period of roughly 200 Myr \citep{2009A&A...507..661B}. The correlation reflects the structural self-similarity found in cluster properties, which is due to the self-similarity of the underlying DM halo and its gravitiational potential \citep{1996ApJ...462..563N}.
\end{enumerate}

\section{Hadronic Models}\label{sect.hadr}
Relativistic protons ($p/m_\mathrm{p} c > 1 \,\mathrm{GeV}$) might be injected in the ICM alongside CRe. CRp have long life-times and are trapped in the cluster volume \citep{1997ApJ...487..529B,1999APh....11...73V}. If their coupling to turbulence  is inefficient their spectrum may only be slightly modified during their life-time so they are assumed to retain their power-law injection spectrum \citep{1999APh....12..169B}: 
\begin{align}\label{eq.CRpSpectrum}
    n_\mathrm{p}(p) \,dp &= K_\mathrm{p}p^{-\alpha_{\mathrm{CRp}}}\,dp,
\end{align}
with a spectral index $\alpha_\mathrm{p}$. From accelerator data we know that CRp inject secondary CRe at GeV energies and $\gamma$-rays during collisions via a hadronic cascade \citep{1986A&A...157..223D}:
\begin{align*}
	p_{\mathrm{CR}} + p_{\mathrm{th}} &\Rightarrow \pi^{0}+\pi^{+}+\pi^{-} + \mathrm{anything}\\ 
    \pi^{\pm} &\Rightarrow \mu + \nu_{\mu}  \\ 
    \mu^{\pm} &\Rightarrow e^{\pm} + \nu_{\mu} + \nu_{e} \\  
    \pi^{0}	 &\Rightarrow  2\gamma.
\end{align*}
This allows to calculate the CR electron spectrum $n_\mathrm{e}(p)$ directly from the thermal density and the injection $Q({\bf r},  p', t)$ of CRe from the CRp spectrum \ref{eq.CRpSpectrum}. Under the assumption of slow varying MHD conditions in the ICM, injection and losses (eqs. \ref{eq.radlosses}+\ref{eq.ilosses}) of CRe balance, the spectrum equillibrates and becomes stationary. The CR transport equation \ref{eqn.fkp} can be simplified to: 
\begin{align}
	\frac{\partial}{\partial p} \left( n_\mathrm{e}(p) \left.\frac{\mathrm{d}p}{\mathrm{d}t}\right|_{\mathrm{loss}}  \right) &= Q({\bf r}, p, t) \label{eq.stationfkp}
\end{align}
The solution of this equation is then trivial \citep{2000A&A...362..151D}:
\begin{align}
    n_\mathrm{e}(p) &= \left|\frac{\mathrm{d}p}{\mathrm{d}t}\right|_{\mathrm{loss}}^{-1} \int\limits^{\infty}_{p} \,\mathrm{d}p'\, Q({\bf r},  p', t). \label{eq.hadsol}
\end{align}
Then the synchrotron emissivity follows \emph{analytically} from the CRe for power-law spectra \citep{1965ARA&A...3..297G,1994hea2.book.....L,1986rpa..book.....R,1970ranp.book.....P}. For a spectrum of CRp like equation \ref{eq.CRpSpectrum} the synchrotron emissivity scales like \citep[e.g.][]{2000A&A...362..151D}:
\begin{align}
    j_\nu &\propto n_\mathrm{th} K_\mathrm{p} \nu^{-\alpha_\nu} \frac{B^{\alpha_\nu+1}}{B^2+B_\mathrm{CMB}^2}, \label{eq.hadrScaling}
\end{align}
with\footnote{when considering the Thomson pp cross-section. The full energy dependent cross-section gives a slightly different scaling. Given the current observational data the difference can be ignored.} $\alpha_\nu = (\alpha_\mathrm{e}-1)/2 = (2\alpha_\mathrm{p}-1)/3$ the spectral index of the radio synchrotron emission and $B_\mathrm{CMB}/(1+z)^2 = 3.2\,\mu\mathrm{G}$ the magnetic field equivalent of the CMB induced IC effect \citep{1994hea2.book.....L}. More elaborate models can be constructed which are based on a piecewise solution of the diffusion equation of CRp \citep{2001ApJ...562..233M} or adiabatic invariants \citep{2007A&A...473...41E}. This has also been combined with a prescription for primary CRe and CRp injection in shocks (i.e. models for radio relics) \citep{2008MNRAS.385.1211P}. Most models share the simplification of \emph{stationarity of the CRe spectrum} obtained from the (sometimes non-stationary) CRp spectrum. In that case CR spectra and subsequent synchrotron spectra are limited to power-laws, sometimes in a number of energy bins. The coupling to turbulence is neglected in these works.\par
In hadronic models the injection efficiency for CRp and CRe in shocks is a free parameter. The $\gamma$-ray emissivity can be computed analytically as well \citep{2004A&A...413...17P}. \par
The magnetic field in clusters is then often modelled relative to the thermal density \citep{2001A&A...378..777D,2010A&A...513A..30B}:
\begin{align}
	B(r) &= B(0) \left( \frac{\rho(r)}{\rho(0)} \right)^{\eta_\mathrm{mag}}.
\end{align}
Free model parameters are then:
\begin{enumerate}
	\item $K_\mathrm{p}(r) = X_\mathrm{CRp}(r) n_\mathrm{th}(r)$: the CRp normalisation is often given relative to the thermal density ($X_\mathrm{CRp} \in [10^{-5},0.1]$) or simulated directly in a numerical approach. It is constrained by the synchrotron profile and has to be consistent with the observed non-thermal pressure and $\gamma$-ray limits.    
    \item $\alpha_\mathrm{p} \in ]2.0,3.0[$: the CRp spectral index translates to CRe and synchrotron spectal index and is therefore constrained by observed radio halo spectra. The spectral index enters the $\gamma$-ray emissivity as well. The observed break in the radio spectrum of Coma is often (falsely) attributed to the SZ-decrement. This is then combined with a flatter $\alpha_\mathrm{p}$ and a flat magnetic field to fit $\gamma$-ray constrains \citep{2012ApJ...757..123A}. We will discuss this in section \ref{had.sz}. 
		\item $B(0) \in [1, 10] \, \mu\mathrm{G}$: the cluster magnetic field normalisation is directly constrained from RM observations, equipartition arguments and simulations.
	\item $\eta_\mathrm{mag} \in [0,1]$: the cluster magnetic field scaling relative to the thermal density. While being constrained by the same observations as the normalisation some models assume (nearly) flat magnetic fields.
\end{enumerate}

\subsection{Simulations \& CR formalism}
	Exemplary we show here results from constrained cosmological MHD simulations published in \citet{2010MNRAS.407.1565D,2010MNRAS.401...47D}. These are consistent with analytic comparisons with newest data \citep{2012MNRAS.426..956B}. \par
The constrained power-spectrum initial conditions produces matter structures which on scales of a few Mpc (clusters) can be identified with their real counterparts \citep{2002MNRAS.333..739M}. The magnetic field was initialised from galactic outflows as shown in \citet{2009MNRAS.392.1008D}. The radio emission is then estimated with an analytic approach to the CR physics: we use the high energy approximation to the p-p cross-section in \citet{2005MNRAS.363.1173B} to compute the synchrotron emissivity of a cluster. We estimate the $\gamma$-ray brightness from our models using the formalism by \citet{2004A&A...413...17P}.\par
\begin{figure}
	\includegraphics[width=0.45\textwidth]{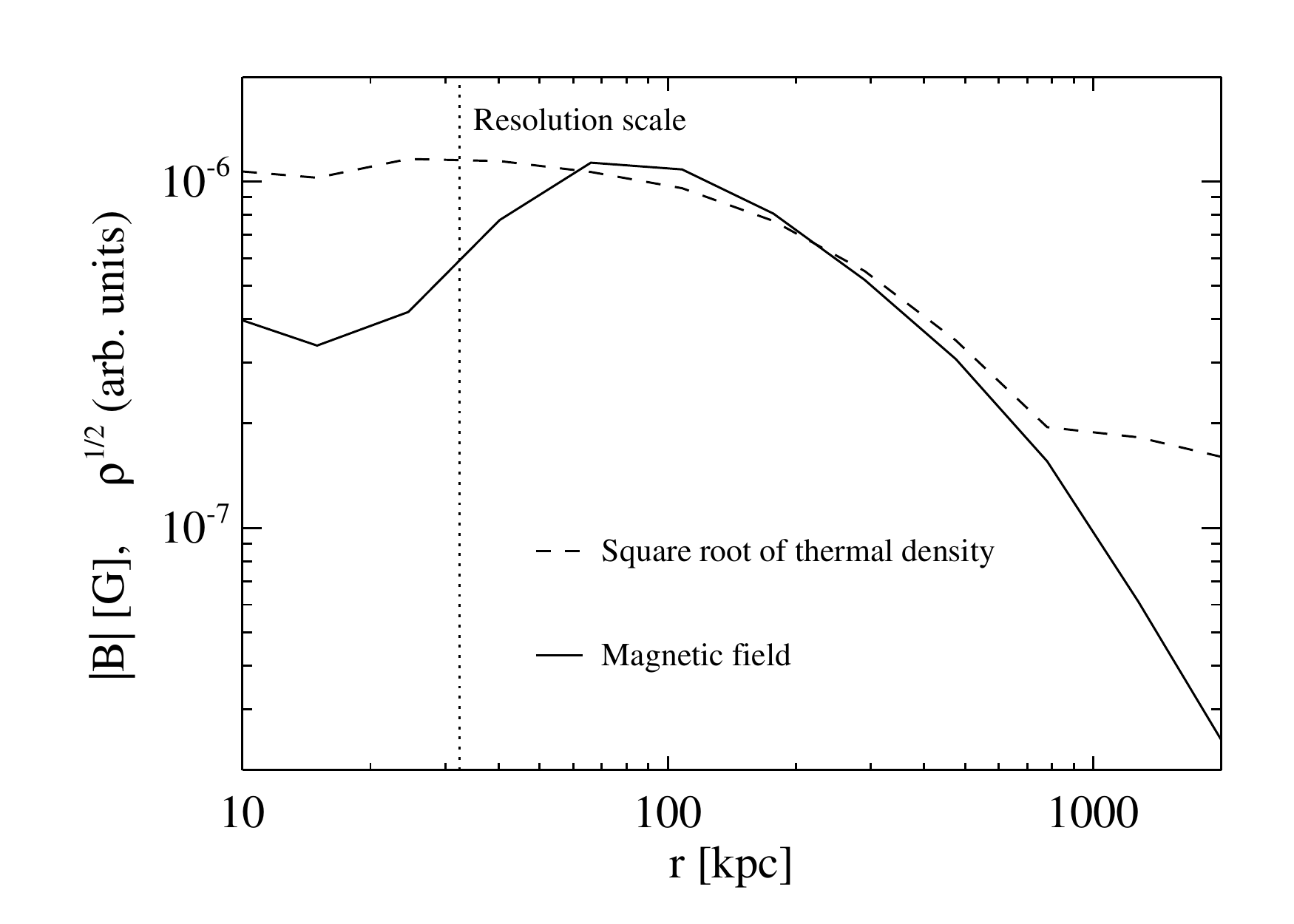}
	\caption{Radial profile of magnetic field strength and square root of the density (dashed) of the simulated cluster. The smallest resolution scale $2h_\mathrm{sml}$ is marked as dotted line.}\label{img.bfld}
\end{figure}

We model the radial profile of CRp normalisation $X_\mathrm{CRp}(r)$ relative to the thermal density as: 
\begin{align}
	X_\mathrm{CRp}(r) &= X_\mathrm{CRp}(0) \left( \frac{\rho(r)}{\rho(0)} \right)^{-\eta_\mathrm{CRp}}.\label{eq.hadscal}
\end{align}
We will fix the normalisation $X_\mathrm{CRp}(0)$ and the scaling $\eta_\mathrm{CRp}$ in the next section to fit the observed radio profile. The proton spectral index $\alpha_\mathrm{CRp}$ is another free parameter which we will constrain from the observed spectrum. The magnetic field is taken from the simulation and is roughly consistent with observations in Coma with $\eta = 0.5$ and $B_0 \approx 1 \,\mu\mathrm{G}$ \citep{2010A&A...513A..30B}. The result is shown in figure \ref{img.bfld}, where we the magnetic field profile is plotted alongside the square root of the density.  \par

\subsection{Results}
In this section we will focus on the Coma cluster first. We show how multi-frequency observations constrain the hadronic model. An overview of the parameters used can be found in table \ref{tbl.param}. Then we show the P14-LX correlation for a sample of simulated galaxy clusters and comment on the implications of the PLANCK correlation between thermal pressure and radio emission in Coma. 
\begin{figure*}
	\centering
	\includegraphics[width=0.495\textwidth]{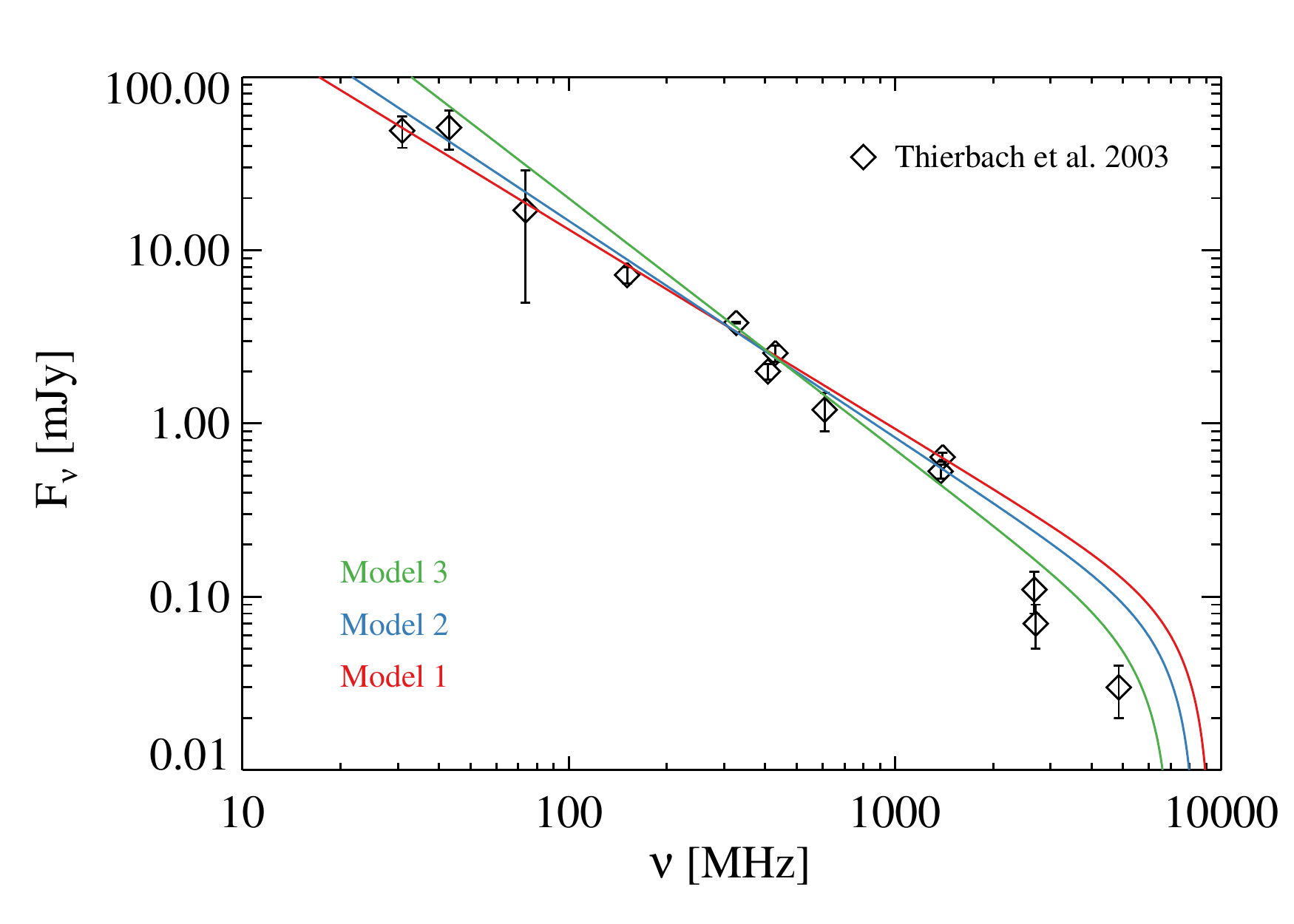}
	\includegraphics[width=0.495\textwidth]{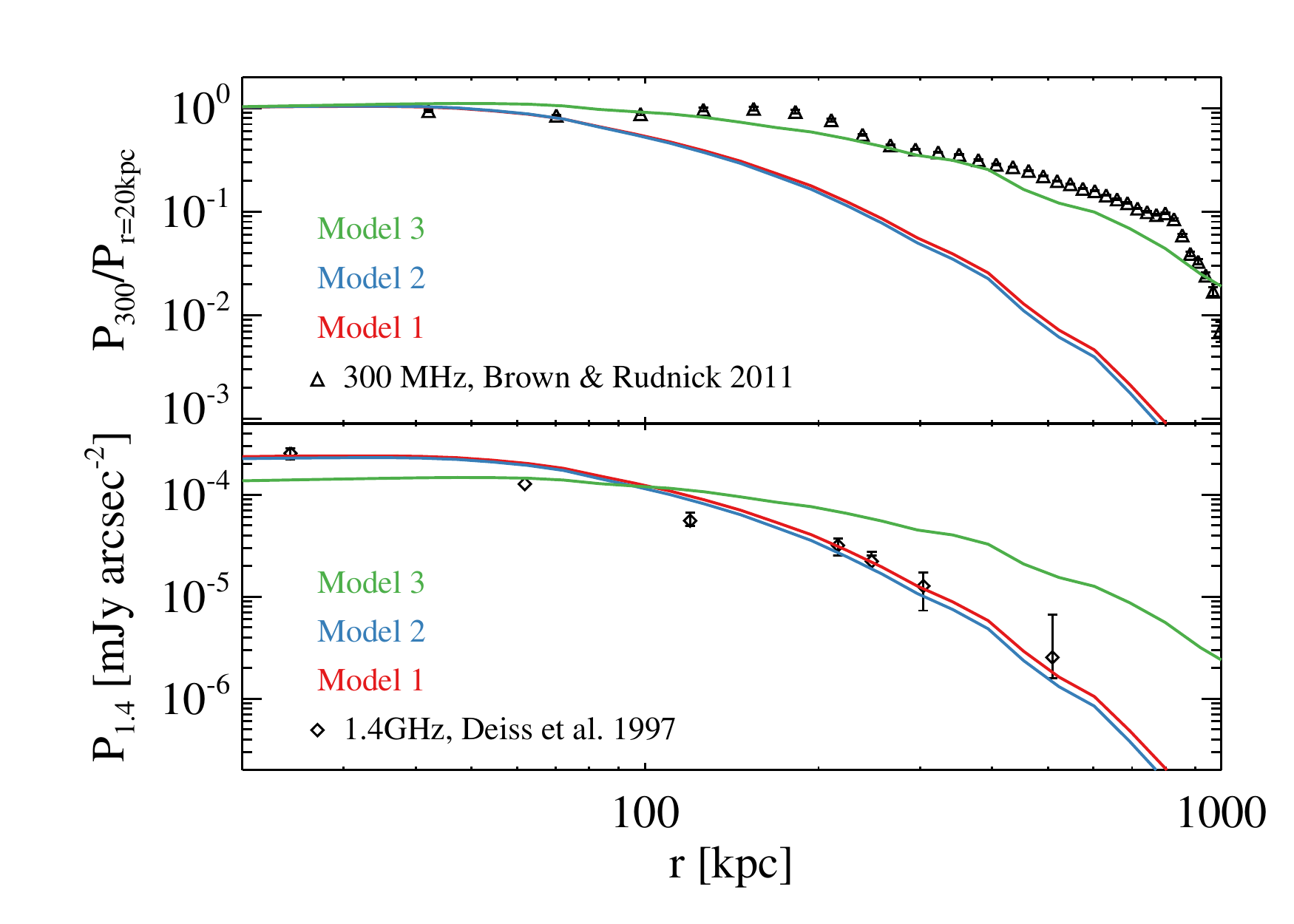}\\
	\includegraphics[width=0.495\textwidth]{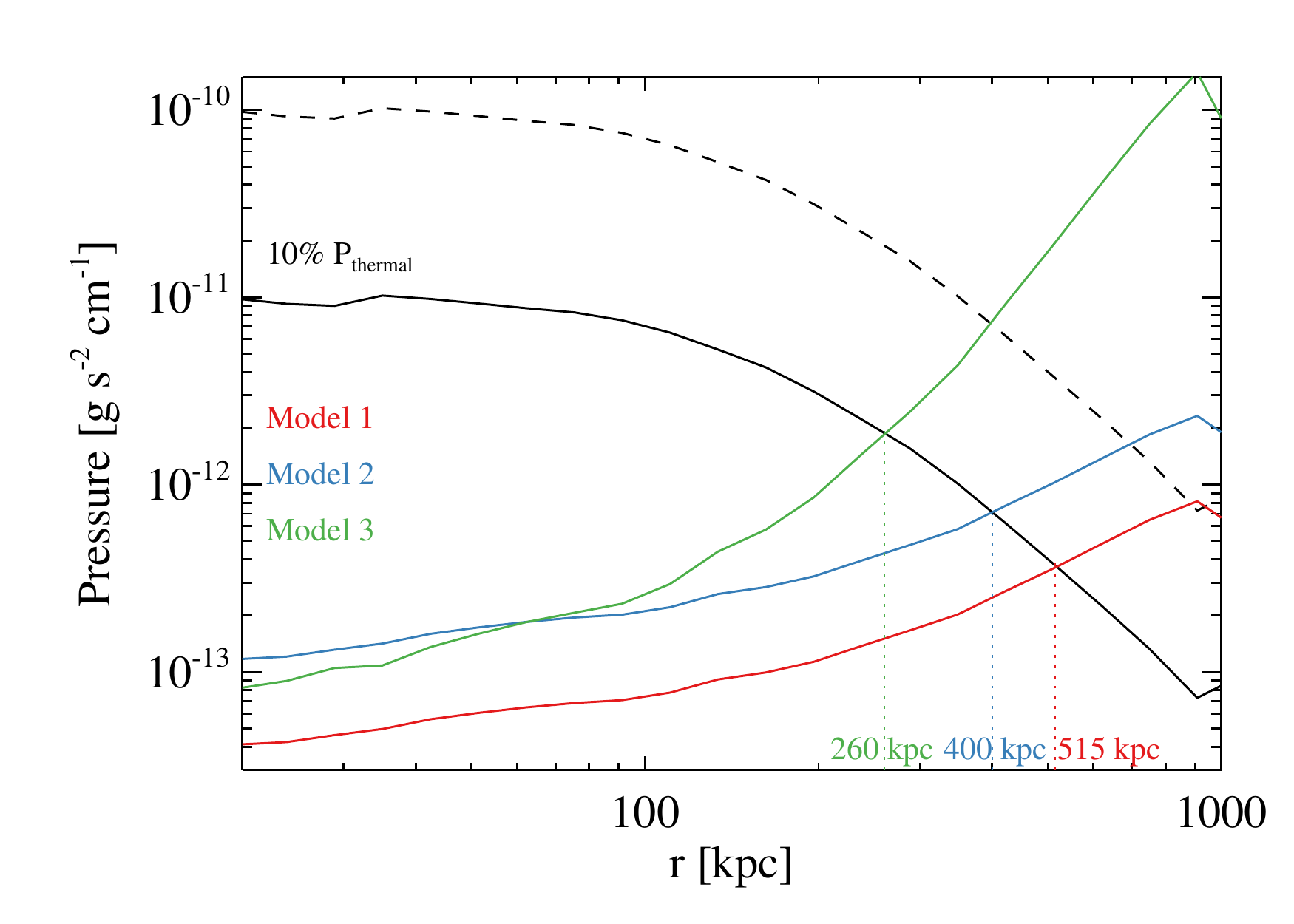}
	\includegraphics[width=0.495\textwidth]{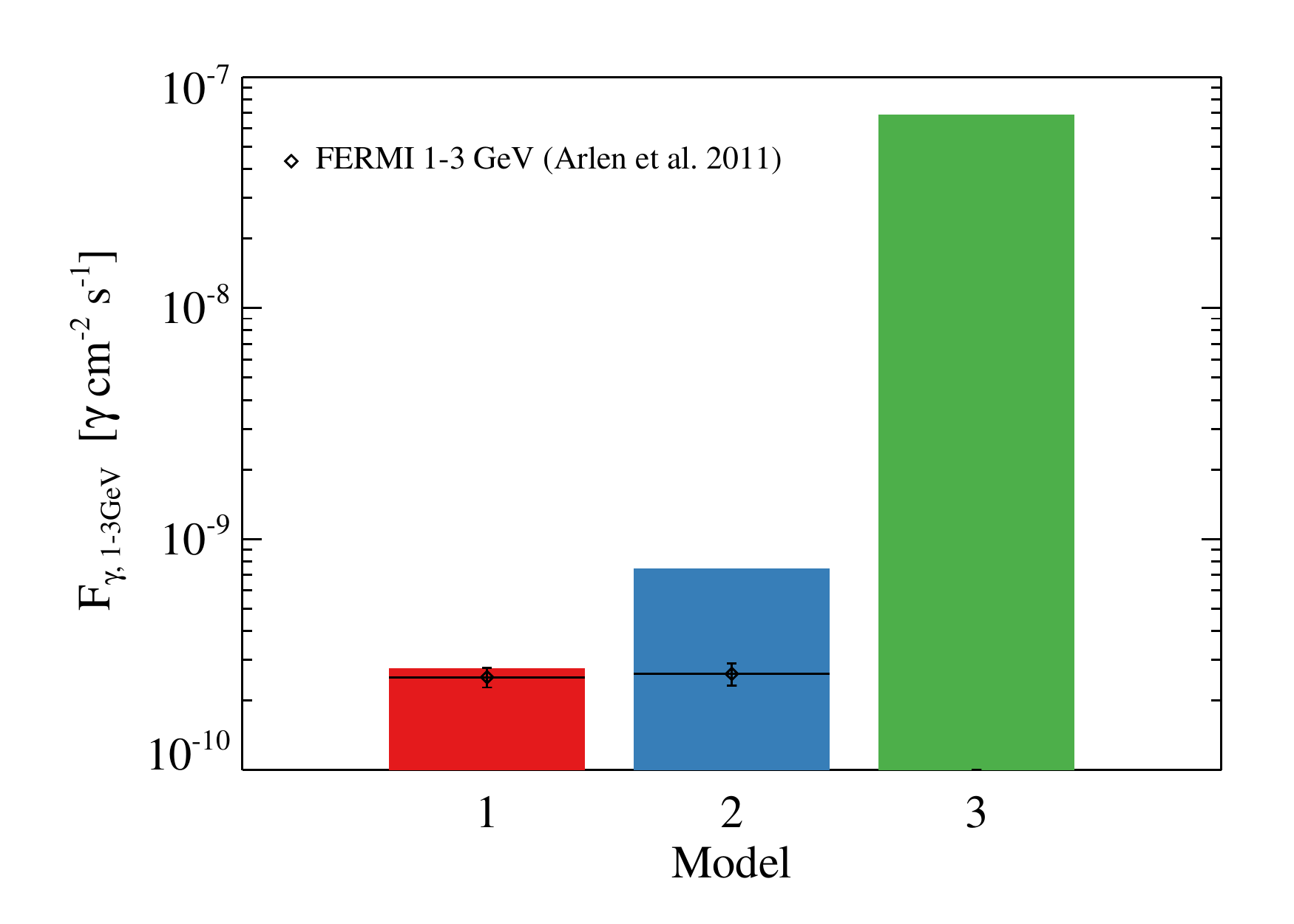}
	\caption{Top left: Observed radio spectrum in the Coma cluster, black
	diamonds \citep{2003A&A...397...53T}. Spectra of three hadronic models with $\alpha_\mathrm{CRp} = 2.3, 2.6, 2.9$ and $\eta_\mathrm{CRp} = 1.4, 1.4, 2.4$. The SZ-decrement has been included pixel by pixel from the observed PLANCK map (Brunetti et al. in prep.) . Top right: Radial profile of radio synchrotron emission in Coma at 300 MHz \citep[top, normalised, ][]{2011MNRAS.412....2B} and 1.4 GHz \citep[bottom, ][]{1997A&A...321...55D}. In colors the three models with a CRp scaling to fit the conservative profile at 1.4GHz (red, blue) and the deep profile at 300 MHz (green). Bottom left: Non-thermal pressure constraint in Coma \citep[black line,][]{2012MNRAS.421.1123C} and non-thermal pressure from the numerical models. Bottom right: Fermi upper limits  at $\alpha_\mathrm{CRp} = 2.3, 3.5$ from \citet{2012ApJ...757..123A} and expected $\gamma$-ray flux from the three numerical models.}
	\label{img.hadrResults}
\end{figure*}

\subsubsection{Coma: Spectrum \& SZ-decrement} \label{had.sz}
	Hadronic models predict a power-law synchrotron spectrum (equation \ref{eq.hadrScaling}), in contrast to the observed break in the Coma spectrum. It has been argued that the break can be explained by the Sunyaev-Zeldovich decrement in the unavoidable CMB radio signal measured alongside the actual halo emission \citep{2002A&A...396L..17E,2004A&A...413...17P}. The SZ-decrement is the modification of the CMB thermal emission by the clusters Sunyaev-Zeldovich effect  at halo frequencies \citep{1980ARA&A..18..537S}. This is not connected to the halo emission, but contaminates the total emission measured from the halo. Naturally the influence is limited to the region of the radio halo only. However \citet{2002A&A...396L..17E} considered the flux from a region more than twice the size of the radio halo (5 Mpc radius), therefore overestimating the decrement. \par
Due to the high resolution of the PLANCK measurements, the SZ-decrement can now be directly estimated from observations. It is instructive to assume three spectral indices: $\alpha_\mathrm{CRp} = 2.3$ and $\alpha_\mathrm{CRp} = 2.6$, as well as $\alpha_\mathrm{CRp} = 2.9$.  In figure \ref{img.hadrResults}, top left we show the Coma radio spectrum and three analytical hadronic models with these spectral indexes, including the SZ-decrement from the PLANCK data (preliminary results from Brunetti in prep.). The spectra are normalised to 300 MHz, where the influence of the decrement is negligible. Only the last model leads to a marginally consistent fit of the observed radio spectrum. A hadronic model with spectral index of 2.6 is already more the a factor of two away from the observations. Therefore even given the uncertainties in the observations (see section \ref{sect.obs}) the SZ-decrement is not able to reproduce the break for indizes smaller than 2.6. 

\begin{table}
	\centering
	\begin{tabular}{c|c|c|c}
		Parameter & Model 1 & Model 2 & Model 3 \\\hline
		$\alpha_\mathrm{CRp}$ & 2.3 & 2.6 & 2.9 \\
		$\eta_\mathrm{CRp}$ & 1.4 & 1.4 & 2.4\\
		$X_\mathrm{CRp}(0)$ & 0.002& 0.006 & 0.006 \\ 
		$B_0 [\mu\mathrm{G}]$ & 1 & 1 & 1 \\
		$\eta_\mathrm{mag}$ & 0.5 & 0.5 & 0.5 \\
		fit $\nu \,\mathrm{MHz}$ & 1400 & 1400 & 300 \\
	\end{tabular}
	\caption{Parameters and values for the three numerical models used in this comparison. }\label{tbl.param}
\end{table}

\subsubsection{Coma: Radial Profiles}
The total radio flux and radial profile of the emission in Coma can be used to 	constrain the amount of CRp in the cluster, i.e. $X_\mathrm{CRp}(r)$. We set the scaling according to equation \ref{eq.hadscal} to fit (table \ref{tbl.param}) the observed radio profile at 1.4 GHz  for model one and two \citep{1997A&A...321...55D}, and 300 MHz for model three \citep{2011MNRAS.412....2B}. The fits are shown for all three models in figure \ref{img.hadrResults}, top right graph. This implies $\eta_\mathrm{CRp} = 1.4$ for the models with small spectral index and $\eta_\mathrm{CRp} = 2.4$ for the other one. The total luminosity of the models is fixed at 1.4 GHz to be $640\pm5 \,\mathrm{mJy}/\mathrm{arcsec}^2$ following \citet{1997A&A...321...55D}. \par
The first two models use the older radio profile at 1.4GHz. The is very \emph{conservative} estimates as at low frequencies the halo is observed to be much flatter and larger \citep{1990AJ.....99.1381V,2001A&A...376..803G,2011MNRAS.412....2B}. Indeed the third model uses the state-of-the-art radio observation of the COMA cluster and models the best constrains available on the radio halo problem today.  \par
CRp exert non-thermal pressure to the ICM, similar to turbulence and magnetic fields. Recent observational studies on Coma set an upper limit to non-thermal pressure of 10 \% in the cluster \citep{2012MNRAS.421.1123C}. For comparison we plot in figure \ref{img.hadrResults} bottom left, the thermal pressure profile (dashed line) alongside the 10\% limit (full line) and the non-thermal pressure of the three models (red, green, blue). The pressure constrain is violated at radii of $r_{10} \approx 510, 400, 260 \,\mathrm{kpc}$, while the observed radio halo at 300 MHz extends to $\approx 1000 \, \mathrm{kpc}$. Therefore observations imply considerable non-thermal pressure at large radii for hadronic models. For the Coma cluster $r_{10}$ could be increased by flatter magnetic field models for flat CRp spectra. However this becomes less consistent with magnetic fields derived from observed rotation measures. \par
The situation becomes increasingly problematic when considering halos like A512. The steep spectrum of these objects require abundant CRp contents to reproduce the emission at lowest frequencies. Observed synchrotron spectra of $\alpha_\nu \approx 2.1$ imply $\alpha_\mathrm{CRp} = 3.5$. This results in non-thermal pressures of more than 50 \% within 3 core radii \citep{2008Natur.455..944B,2009ApJ...699.1288D}.\par

\subsubsection{$\mathbf{\gamma}$-ray brightness}
Observations of $\gamma$-ray emission from hadronic collisions in the ICM are an independent way of constraining the CRp content in clusters. Newest upper limits in this frequency band come from FERMI and VERITAS, while the former ones are most constraining \citep{2012ApJ...757..123A}. \par
In figure \ref{img.hadrResults} bottom right, we show the most constraining upper limits from FERMI ($\alpha_\mathrm{CRp} = 2.3, 2.5$) as well as our three numerical models. Only the first model with the flattest CRp spectrum is marginally consistent with the newest $\gamma$-limits given the simulated magnetic fields. The other two steeper models violate the observed limits. Again the situation could be eased by a stronger and flatter magnetic field profile in the case of flat CRp spectra. The steep $\alpha_\mathrm{CRp}$ model however can be considered excluded by $\gamma$-ray observations. It is the only one roughly consistent with the observed radio spectrum though. 

\subsubsection{Radio - X-ray Correlation}
\begin{figure*}
	\centering
	\includegraphics[width=0.9\textwidth]{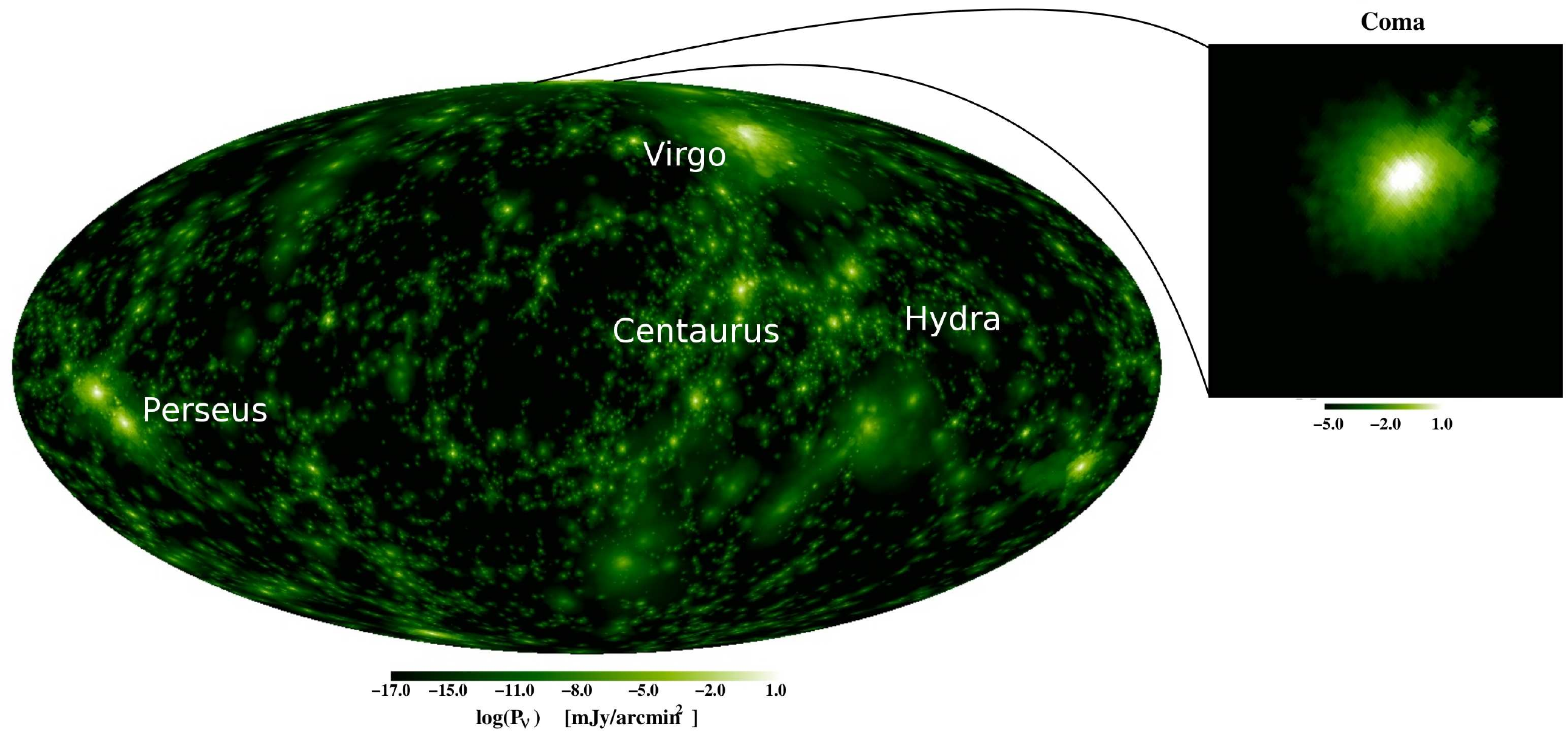}
	\caption{Full sky map of radio flux from a hadronic model with $X_\mathrm{CRp}(r) = 1\%$ and $\alpha_\mathrm{CRp} = 2.6$ in a cosmological MHD simulation.}
	\label{img.fullsky}
\end{figure*}
\begin{figure}
	\centering
	\includegraphics[width=0.45\textwidth]{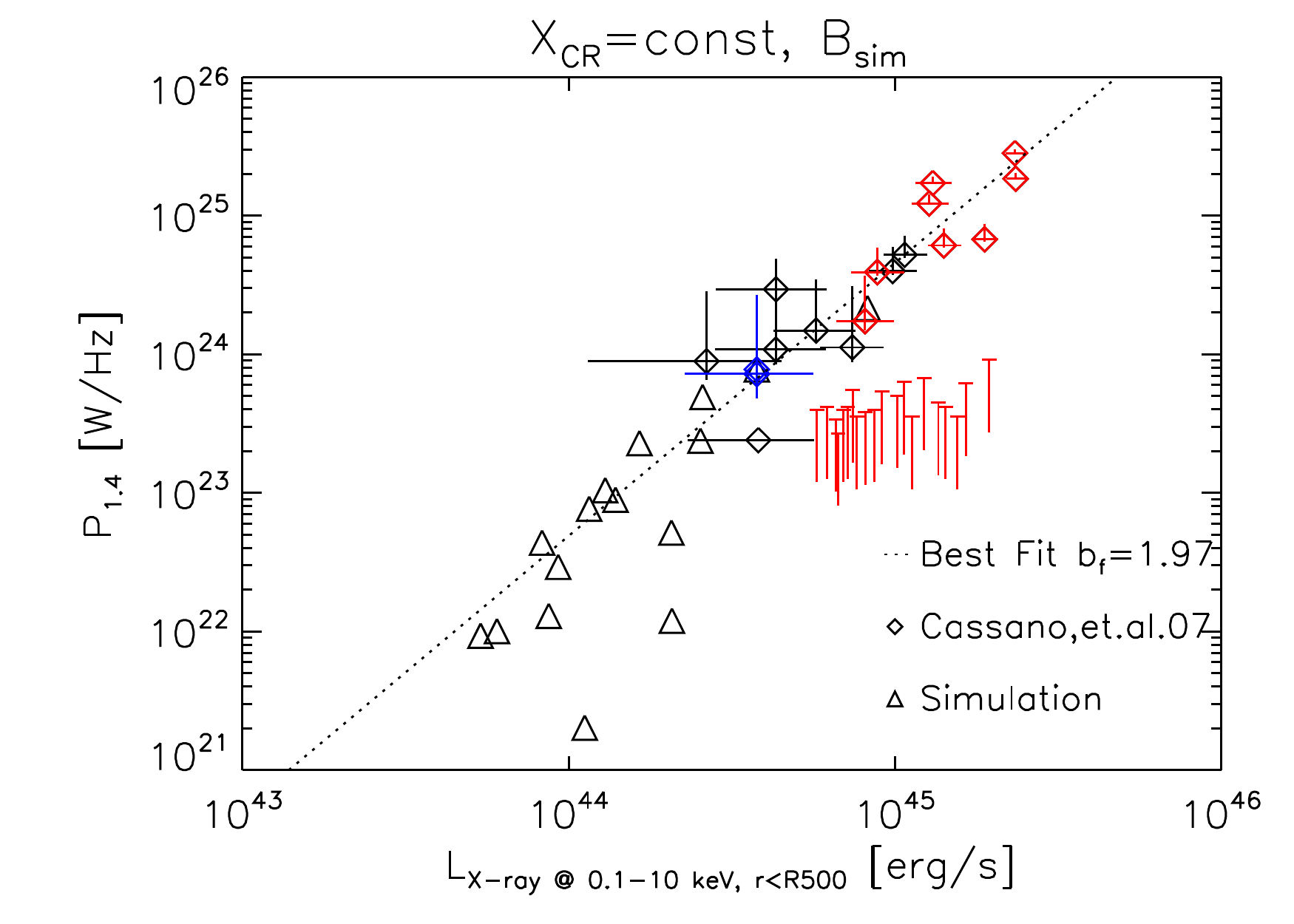}
	\caption{Radio brightness at 1.4GHz over X-ray brightness between 0.2 and
		2.4 keV for 16 clusters with $X_\mathrm{CRp}(r) = 1\%$ and $\alpha_\mathrm{CRp} = 2.6$. Observed halos as diamonds from \citep{2007MNRAS.378.1565C} and upper limits from \citep{2007A&A...463..937V}. }
	\label{img.correlation}
\end{figure}
Figure \ref{img.correlation} shows 16 simulated clusters from the cosmological simulation in the $P_{14\mathrm{GHz}} - L_\mathrm{X-ray}$ plane (triangles, $X_\mathrm{CRp}= 1\%$, $\alpha_\mathrm{CRp} = 2.6$). Overplotted are the observed correlation (dotted line), observed radio halos (red \& white diamonds, \citet{2007A&A...463..937V}) and upper limits (red lines). The observed correlation is well reproduced by the largest clusters in the simulated sample. \par
However the brightness distribution from the simulation is not bimodal: all large clusters show significant radio emission. This is an intrinsic prediction of hadronic models. As CR protons are trapped in the cluster volume they \emph{accumulate} in the ICM. In figure \ref{img.fullsky} we show a full sky projection of the radio synchrotron emission from the simulation using the $X_\mathrm{CRp}=const$ model and the simulated magnetic field. Every cluster shows diffuse radio emission. The radio sky predicted by this model is shown in figure \ref{img.fullsky}. \par

\subsection{Summary}
Given our simulated magnetic fields, hadronic models:
\begin{enumerate}
    \item reproduce the $P_{14\mathrm{GHz}} - L_\mathrm{X-ray}$ correlation, but in the case of classical models fail the bimodal distribution of the population.
    \item do not reproduce the deepest observed radio profile in Coma within the non-thermal pressure constrains. This could be eased using a flatter and stronger magnetic field distribution. However this tends to be not consistent with observed RM measures in the Coma cluster. 
	\item are not consistent with present upper limits in the $\gamma$-ray regime when reproducing deepest Coma radio observations and its radio spectrum. Models using $\alpha_\mathrm{CRp} = 2.3$ are indeed consistent with these limits if the 1.4GHz profile is fitted.  However they do not fit the Coma spectrum (figure \ref{img.hadrResults}, top left).
    \item do not reproduce the break in the Coma spectrum when $\alpha_\mathrm{CRp} < 3$.  
\end{enumerate}
The observed constrains present serious challenges for hadronic models. The observed spectrum of the Coma cluster is impossible to fit without violating non-thermal pressure constrains and $\gamma$-ray limits. In the most recent attempts \citet{2012ApJ...757..123A,2012arXiv1207.6410Z} report a good fit to the observed constrains.  However in these papers Coma is modelled with\footnote{because of the over-estimation of the SZ -decrement} $\alpha_\mathrm{CRp} = 2.1-2.4$ at 1.4 GHz only !

\paragraph{The PLANCK Compton-y - Radio Correlation: }
The situation can be presented most elegantly from the observed Compton-y - 300 MHz radio profile correlation  in Coma \citep{2012AAS...22050705M}. In the center of the cluster, the fit in figure \ref{img.hadrResults} (and most other hadronic models) predict $X_\mathrm{CRp} \approx 10^{-3} - 10^{-2}$ (depending on the CRp spectral index) to reproduce the luminosity profile locally. At the outer radii ($300<r<1000 \, \mathrm{kpc}$), where $B << B_\mathrm{CMB}$, PLANCK found:
\begin{align}
    y_\mathrm{Compton} & \propto j_\mathrm{1.4GHz}(r).
\end{align}
The Compton-y parameter measures pressure, so:
\begin{align}
    y_\mathrm{Compton}(r) &\propto P \propto n_\mathrm{th} T,\\
    j_\mathrm{1.4GHz} &\propto X_\mathrm{CRp} n_\mathrm{th}^2 B^{2}.
\end{align}
The temperature profile in clusters is roughly constant with radius, and we have
\begin{align}
	X_\mathrm{CRp} &\propto n_\mathrm{th}^{-1} B^{-2}, \quad \mathrm{and} \\
	 n_\mathrm{th} &\propto (1+r^2/r_\mathrm{c}^2),
\end{align}
for $r \le r_{500}$. It follows from the PLANCK correlation that $X_\mathrm{CRp} \propto  (1+r^2/r_\mathrm{c}^2)$, even for a flat magnetic field. In Coma $r_\mathrm{c} \approx 0.3 \,r_{500}$, so $X_\mathrm{CRp}$ has to increase by a factor of 10 in this case.  If $\alpha_\mathrm{CRp}>2.9$ then $X_\mathrm{CRP} \approx 0.01$ and the non-thermal pressure constraint of 10\% is violated at 1Mpc even for flat magnetic field models. \par
However observed rotation measures suggest $B \propto n_\mathrm{th}^{1-0.5}$ \citep{2010A&A...513A..30B}. Here the increase is a factor\footnote{This is consistent with our fit (figure \ref{img.hadrResults}, left, cyan curve). } of 100 to 1000 and the non-thermal pressure constraint is violated in any case.  

\subsection{A Comment on Non-classical Hadronic Models}\label{nonclassical}
	Recently \citet{2011A&A...527A..99E} proposed a model based on CRp transport in the ICM in an attempt to explain the bimodal distribution of halos in the hadronic framework. In this model the bimodality reflects two modes of CRp transport in clusters: 
\begin{itemize}
	\item During mergers CRp transport is dominated by convection through turbulent motions. CR protons are effectively stored between converging magnetic field lines\footnote{Similar to magnetic bottles} and dragged along the turbulent motions. These motions are injected at a scale of the core radius and induce an MHD cascade of cluster-wide turbulence. This way the CRp density in the center of clusters is increased. The resulting flatter $X_\mathrm{CRp}$ profile leads to the observed radio bright clusters. 
	\item After turbulence has decayed CRp transport is supposed to be dominated by super-Alvenic streaming of CRp along the magnetic field lines.  If small scale turbulence is not driven by external processes the streaming instability is the only source of fluctuations at that scale. Scattering of CRs on these fluctuations limits the diffusion speed to the Alven speed \citep{1974ARA&A..12...71W}. Specifically the isotropisation of CRp for small pitch angles (i.e. reversal of direction along the field) is mediated by scattering with large modes (i.e. mirror interactions) \citep{2001ApJ...553..198F}. However these modes are damped efficiently by the ICM plasma through the ion cycloton resonance. \citet{2011A&A...527A..99E,1979ApJ...228..576H} argue that this way the streaming instablity is inefficient to generate modes at small pitch angles. This would allow streaming motions along the field lines comparable to the sound speed (highly super-Alvenic). Eventually this tends to remove CRp from the cluster center and flatten the CRp profile of the cluster. Subsequently the radio brightness declines. This process might as well be able to explain the steepening of the radio spectrum as a superposition of different CRp populations.
\end{itemize}
However the assumption of streaming velocities higher than the Alven speed had been rejected earlier for CRe and only lead to the cooling time dilemma. We therefore see three problems in the argumentation presented in \citep{2011A&A...527A..99E,1979ApJ...228..576H} and above: 
\begin{figure}
	\centering
	\includegraphics[width=0.45\textwidth]{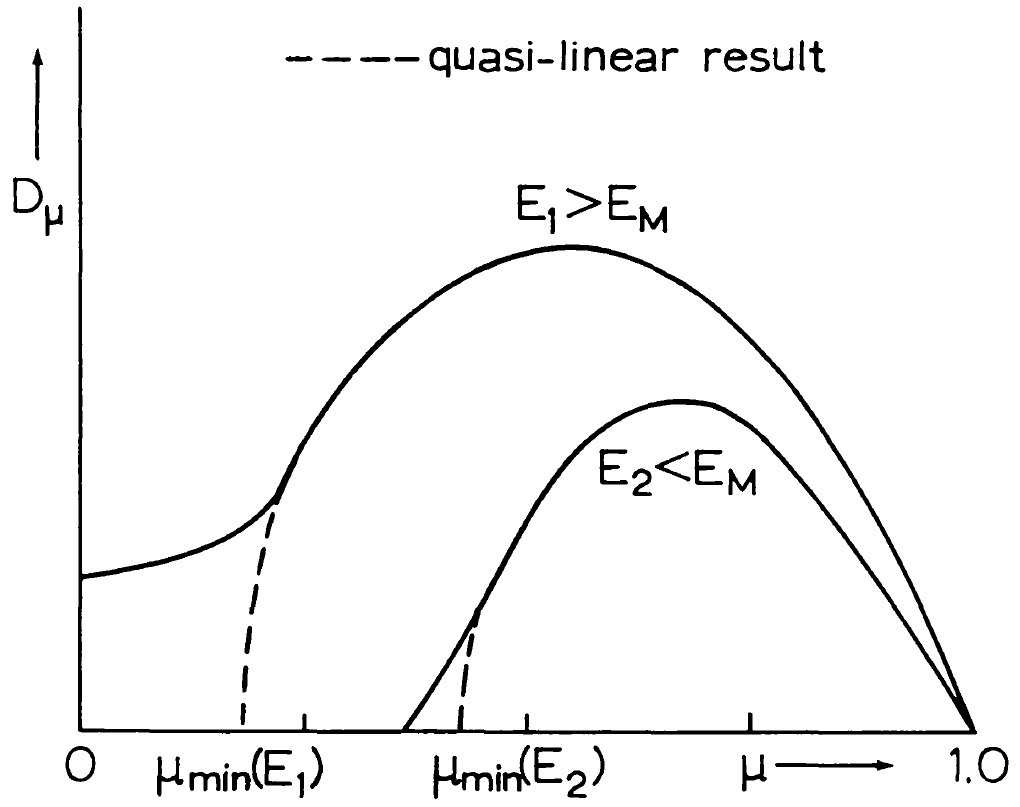}
	\caption{Diffusion coefficient over pitch angle in radians \citep{1981A&A....98..161A}. The quasi-linear result is shown dashed, the non-linear result as full curve. In clusters $E_1 > E_\mathrm{M}$.} \label{img.pascattering}
\end{figure}

\begin{itemize}
    \item  Cosmological simulations show that even in relaxed clusters, structure formation causes a constant infall of small halos into the ICM, driving turbulence. The ICM itself is not a classical collisional plasma. Probably collective effects mediate collisionality, which implies very high Reynolds numbers of the plasma \citep{2011MNRAS.412..817B}. Therefore the infall of structures always causes turbulence on kpc scales which is dissipated at sub-pc scales. This is in contrast to the non-classical hadronic approach which relies on small modes generated exclusively by the streaming instability. Even though damping processes are ill constrained in the ICM it is likely that externally driven turbulence is always present at some level on the damping scale of CRp.
    \item \citet{1979ApJ...228..576H} argue that at pitch angles of $\mu < 2\,\mathrm{deg}$ turbulence is completely absent due to the strong damping by the thermal ions. However even in the absense of external driving, turbulence is constantly injected on a range of scales by CRp through the streaming instability. Due to mode coupling these motions will form a cascade and develop \emph{a break with finite steepness} at the wavelength of interest \citep[][]{1986PhFl...29.2535S,2002cra..book.....S}. Therefore turbulence is unlikely to be completely absent, even if ion cycloton damping is stronger than the growth rate from the streaming instability.
    \item \citet{1979ApJ...228..576H} calculate the diffusion coefficient in the quasi linear regime. In that formalism the resonance function $R_j(\vec{k}, \omega_j)$ is approximated by a delta function \citep[e.g.][]{2002cra..book.....S}:
    \begin{align}
        R_j({\bf k},\omega_j) &= \pi \delta(v\mu k_\parallel - \omega_\mathrm{R,j} + n\Omega). \label{eqn.qsRapprox}
    \end{align}
    However, this approximation is only valid if $\Gamma_j \rightarrow \infty$, i.e. \emph{damping of plasma modes is negligible} \citep{1980panp.book.....M}. This is not the case in the situation considered here. Ion cycloton damping obviously has to be considered, because it is supposed to dominate at small pitch angles. The correct resonance function is of Breit-Wigner type \citep[e.g.][]{2002cra..book.....S}:
    \begin{align}
        R_j({\bf k},\omega_j) &= \frac{\Gamma_j({\bf k})}{\Gamma^2_j({\bf k}) + (v\mu k_\parallel - \omega_\mathrm{R,j} + n\Omega)}
    \end{align}
    Considering this resonance function the diffusion coefficient has been investigated in non-linear theoretical approaches \citep{1966PhFl....9.1773D,1970PhFl...13.2308W,1969PhFl...12.1045W,1973Ap&SS..25..471V,1975Ap&SS..38..125B,1976ApJ...204..900G,1981A&A....98..161A,1978PhFl...21..347J,2008ApJ...673..942Y}. The resulting diffusion coefficient ($D_\mu$)  is consistent between authors. We show in figure \ref{img.pascattering} the result from \citet{1981A&A....98..161A} which does not vanish for $\mu \rightarrow 0$, in the case of fully relativistic particles. CR scattering through $\mu = 0$ appears not to be a problem in the correct non-linear approach.
\end{itemize}
We conclude that CR streaming is not important in galaxy clusters. 

\section{Reacceleration Models}\label{sect.reacc}
    The ICM thermal plasma might be dominated by field-particle interactions to establish the observed collisionality of the plasma. This can be seen when comparing particle-particle and particle-field interactions: In a cluster, particle collision lengths are $>1 \,\mathrm{kpc}$ and the speed of a particle in the collisionless regime is the sound speed, $\approx 10^8\,\mathrm{cm}/\mathrm{s}$. However the debye length\footnote{the distance it takes to shield the field of an extra charge} is of the order of $20 \,\mathrm{km}$ \citep{2002cra..book.....S}. That means every charged particle is coupled to $\approx 10^6$ neighbouring particles through its em-field, with quasi instantaneous interaction speed (the speed of light). In contrast the time between collisions is $10^{21} \,\mathrm{cm} / 3\times10^8 \mathrm{cm}/\mathrm{s} \approx 1\,\mathrm{Myr}$. \par
    One may consider, that major mergers drag $10^{62}\,\mathrm{erg}$ of kinetic energy into the ICM. The most part is dissipated through shocks into heat, which is prominently seen in radio relics \citep[e.g.][]{2009A&A...504...33V,2012MNRAS.421.3375V}. Additionally, vortical motions, shear flows and instabilities efficiently drive turbulence in the ICM \citep{2006MNRAS.366.1437S}, because of its high Reynolds number. This amplifies magnetic fields to $\mu G$ values \citep{2001A&A...378..777D,2009MNRAS.392.1008D,2012MNRAS.422.2152B}. Furthermore, turbulent modes (local field fluctuations) resonantly couple to the cosmic-ray electrons in the ICM on pc scales \citep{2002cra..book.....S}. This way a synchrotron dark, long-lived population of CRe can diffuse throughout a cluster, and is accelerated to synchrotron bright energies by turbulence, on a time-scale shorter than the life-time of radio halos \citep{2001ApJ...557..560P,2001MNRAS.320..365B,1987A&A...182...21S}.  \par

\subsection{Dynamics of the CRe population}
    Models involving reacceleration do not assume stationarity of the CR population. Therefore the analytical formalism based on the stationarity condition, eq. \ref{eq.stationfkp}, can not be used in this framework. Again the population of cosmic-rays is described by its isotropic spectral density $n(p,t)$ in momentum space. The dynamics of this spectrum is determined by the interplay between cooling, due to radiative and Coulomb losses, and acceleration, due to turbulence (figure \ref{img.coolingtime},left). The momentum transport is governed by a Fokker-Planck equation (\ref{eqn.fkp}), in which these processes are realised as coefficients. The loss terms (eqs. \ref{eq.radlosses}, \ref{eq.ilosses}) define a cooling time scale which is given by equation \ref{eqn.cooltime}, and is of the order of $10^8\,\mathrm{yrs}$ for synchrotron bright CRe in the ICM (see section \ref{intro}).\par
The coupling to turbulence through resonant scattering on plasma waves is realised through the $D_\mathrm{pp}$ coefficient (eq. \ref{eq.dpp}), which defines an acceleration time scale \citep[e.g.][]{2005MNRAS.357.1313C}\footnote{They are missing a factor of two in the corresponding formula}:
\begin{align}
    t_\mathrm{acc}^{-1} &= 4 \frac{D_\mathrm{pp}}{p^2}.
\end{align}
Stochastic acceleration is efficient \emph{only}, if this time scale is comparable or smaller than the cooling time scale in the medium. For the ICM this is true, when the local turbulent velocity is of the order of $\approx 300 \,\mathrm{km/s}$ on scales of $50\,\mathrm{kpc}$.

\subsubsection{The transport equation} 
    The change of the spectrum $n(p)$ in momentum space is described by a Fokker-Planck equation. It follows from the relativistic Maxwell-Vlasov system of equations \citep[see][ and references therein]{2002cra..book.....S}. Here one neglects the spatial diffusion of cosmic-rays in the ICM for reasons layed out in section \ref{nonclassical} \citep{1974ARA&A..12...71W}.
\begin{align}
    \frac{\partial n(p,t)}{\partial t} &= \frac{\partial}{\partial p}\left[ n(p,t)\left( \left|\frac{\mathrm{d}p}{\mathrm{d}t}\right|_{\mathrm{rad}} + \left|\frac{\mathrm{d}p}{\mathrm{d}t}\right|_{\mathrm{ion}} -\frac{2}{p} D_{\mathrm{pp}}(p) \right) \right] \nonumber\\
    &+ \frac{\partial}{\partial p}\left[ D_{\mathrm{pp}}(p) \frac{\partial n(p,t)}{\partial p} \right] + Q_{\mathrm{e}}(p,t).\label{eqn.fkp}
\end{align}
Conceptually this is a non-linear momentum diffusion equation. Setting $D_\mathrm{pp}=0$, one recovers the usual diffusion equation with the loss terms as positive definite diffusion coefficient. This leads to diffusion to lower momenta, i.e. cooling. For  $D_\mathrm{pp}\ne 0$ this diffusion coefficient is modified and can become negative, leading to diffusion to higher momenta, i.e. heating. Additionally, the second term on the right hand side describes a non-linear broadening of the distribution. The evolution of a spectrum is shown in figure \ref{img.reaccresults}, left for constant injection of power-law CRe at momenta $p \in [50, 10^5]\,\mathrm{m}_\mathrm{e}\mathrm{c}$.

\paragraph{The injection of CRe} in the Fokker-Planck equation  is described by  an injection function $Q(p,t)$. Possible injection sources include shock acceleration, hadronic processes, reconnection, AGN and galactic outflows \citep{2007MNRAS.375.1471B,2011MmSAI..82..636L}. In general these processes are found to inject power-law distributions of CRe into the ICM \citep[see e.g.][ for details]{2010MNRAS.402.2807B}.

\paragraph{The loss terms} due to inverse Compton scattering with CMB photons and synchrotron radiation and Coulomb losses at $z=0$ take the form \citep{2005MNRAS.357.1313C}:
\begin{align}
    \left.\frac{\mathrm{d}p}{\mathrm{d}t}\right|_{\mathrm{rad}} &= -4.8\times10^{-4} p^2 \left[ \left( \frac{B_{\mu G}}{3.2} \right)^2 + 1\right],\label{eq.radlosses}\\
    \left.\frac{\mathrm{d}p}{\mathrm{d}t}\right|_{\mathrm{ion}} &= -3.3\times10^{-29}n_{\mathrm{th}}\left[ 1 + \mathrm{ln}\left(\frac{\gamma}{n_{\mathrm{th}}}\right)/ 75 \right],\label{eq.ilosses}
\end{align}
where $B_{\mu\mathrm{G}}$ is the magnetic field in micro Gauss, $n_\mathrm{th}$ the number density of thermal particles and $\gamma$ the Lorentz factor. \par
One may observe that synchrotron emission can be described with an effective magnetic field in the IC formula. This stems from the fact that IC and synchrotron mechanisms use the same quantum-mechanical scattering process \citep{1994hea2.book.....L}. They are dominant at higher momenta ($p > 10^3\,\mathrm{m}_\mathrm{e}\mathrm{c}$). Ionisation losses and Coulomb scattering depend on the thermal number density of particles in the ICM $n_\mathrm{th}$ and are dominant in the low energy regime ($p < 10^2\,\mathrm{m}_\mathrm{e}\mathrm{c}$). 

\paragraph{The momentum diffusion coefficient} follows from a pertubative approach to the relativistic Maxwell-Vlasov system of equations. Here it is assumed that the change of the field perturbation  by the particle can be neglected\footnote{I.e. damping can be neglected; compare with CR streaming, section \ref{nonclassical}, where this is not the case.}. This is referred to as \emph{quasi-linear-theory}. The rigorous derivation is lengthy, see \citet{2002cra..book.....S} for details. We follow here a more physical the approach by \citet{2007MNRAS.378..245B}. \par
The $D_\mathrm{pp}$ coefficient describes the interaction of cosmic-rays with turbulence in a plasma. Turbulent fluctuations manifest in a plasma, amongst others, in fluctuations of the electric and magnetic field.  In the ICM these fluctuations can be damped by cosmic-rays, similar to how a dielectric damps an infalling light-wave. \par
 The \emph{dielectric tensor} describes the action of the medium/CRs on the fluctuating e/m-fields. In the core of this process is the \emph{resonance condition} (eq. \ref{eqn.rescond}), which describes the coupling of the particle population to the e/m-fields. However turbulence is a statistical process and the turbulent fluctuations are better described by a spectrum in $k$-space. This spectrum evolves according to a damped diffusion equation (eq. \ref{eq.turbspec}). To describe the action of the dielectric on the damping of the waves the dielectric tensor has to be translated to a \emph{damping coefficient} (eq. \ref{eqn.damping}).  Using detailed balancing and a Kraichnan spectrum, the diffusion coefficient (eq. \ref{eq.dpp}) can then be derived from an energy argument (eq. \ref{eq.detbal}). \par
\citet{2007MNRAS.378..245B} consider only the scattering by fast compressive turbulent modes\footnote{An MHD version of sound waves}. This is the least efficient coupling, which makes this a \emph{conservative approach}. Alven waves have been considered in e.g \citet{2004MNRAS.350.1174B} and \citet{2005MNRAS.363.1173B}, but require CR protons and their back-reaction on the spectrum. \par
In the presence of compressible low-frequency MHD-waves with an energy spectrum $W(k)$, relativistic particles with velocity $v$ in a magnetic field are accelerated by electric field fluctuations. In quasi-linear theory this gyroresonant interaction  applies for the resonance condition \citep{1968Ap&SS...2..171M}:
\begin{align}
    \omega - k_\parallel v_\parallel -n\frac{\Omega}{\gamma} &= 0, \label{eqn.rescond}
\end{align}
with $\omega$ the frequency of the wave, $k_\parallel$ and $v_\parallel$ the wave-vector and particle velocity parallel to the magnetic field, $\Omega$ the Larmor frequency, $\gamma$ the Lorentz factor and $n$ the resonance order. Here only $n=0$ case is considered (Transit time damping), which requires effective pitch-angle isotropisation of the particles by other processes. \par
The resonance condition (\ref{eqn.rescond}) can be used to derive the damping rate $\Gamma$ of turbulence from the dielectric tensor $K_{ij}^\mathrm{a}$  in the limit of long wavelengths. Note that here indeed one is allowed to use the quasi-linear approximation equation \ref{eqn.qsRapprox}, because the damping is very small. The damping rate then follows from the dielectric tensor via \citep{1968Ap&SS...2..171M}: 
\begin{align}
    \Gamma &= -i \left( \frac{E^*_i K_{ij}^\mathrm{a} E_j }{16\pi W(k)} \right)_{\omega_i=0} \omega_\mathrm{r}, \label{eqn.damping}
\end{align}
where $E_i$ are the electric fields, $\omega_\mathrm{r}$ is the real part of the frequency of the fluctuating field, and $\omega_{i}$ its imaginary part. \par
The turbulence spectrum follows  the damped diffusion equation:
\begin{align}
    \frac{\partial W_\mathrm{k}(t)}{\partial t} &= \frac{\partial}{\partial k} \left( D_\mathrm{kk} \frac{\partial W_\mathrm{k}(t)}{\partial k} \right) - \sum\limits_{i=1}^{N} \Gamma^i(k) W_\mathrm{k}(t),\label{eq.turbspec}
\end{align}
with the diffusion coefficient due to mode coupling $ D_\mathrm{kk} = k^2/\tau_\mathrm{s}$, $\tau_\mathrm{s}$ the spectral energy transfer time. Here $i$ sums over all damping processes of turbulence in the plasma. For the high-$\beta$ ICM thermal particles and CRp argueably do not damp fast modes efficiently. Therefore we only consider CRe damping here ($N=1$).\par
The argument of \emph{detailed balancing} then states that the energy lost by the turbulent waves below a scale $l$ due to CRe damping equals the energy change in the relativistic particles. It relates the total energy in a mode $W(k)\,\mathrm{d}k$ to the particle spectrum \citep{1979ApJ...230..373E,1981A&A....98..161A}:
\begin{align}
    \int \mathrm{d}p \, E_\mathrm{CRe} \left( \frac{\partial f_\mathrm{CRe}(p)}{\partial t} \right) = \int \mathrm{d}k\, \Gamma_\mathrm{CRe}(k, \theta) W(k).\label{eq.detbal}
\end{align}
From this argument $D_\mathrm{pp}$ can be found:
\begin{align}
    D_\mathrm{pp} &\approx 9\times 10^{-8} p^2 \frac{v_\mathrm{turb}^4}{c_\mathrm{s}^{2}l} \eta_\mathrm{turb}^2 \label{eq.dpp}
\end{align}
for Kraichnan turbulence $P(k) \propto k^{-3/2}$ and an energy fraction in magneto-sonic waves $\eta_\mathrm{turb}$.\par

\subsection{Model Parameters}
As reacceleration models are non-stationary, all parameters have to be estimated as a function of time. One may model the thermal quantities with the usual isothermal beta model. This leaves the following model parameters:
\begin{itemize}
    \item Magnetic field distribution, i.e. $B_0(t)$ and $\eta_\mathrm{B}(t)$.
    \item Spectral distribution of turbulence $W(t)$: e.g. a power-law with index $\alpha_\mathrm{turb}$, cut-off $k_\mathrm{min}$, $k_\mathrm{max}$ and normalisation $W_0(t)$. For Kraichnan turbulence this translates to eq. \ref{eq.dpp} with the turbulent velocity $v_\mathrm{turb}$ at scale $l$.
    \item The fraction of turbulence in magneto-sonic waves $\eta_\mathrm{turb} \in [0.01,0.5]$. This parameter is not well constrained, but given the theoretical uncertainties in the coupling (see the motivation of $D_\mathrm{pp}$ above) values between 0.01 and 0.5 seem acceptable.
    \item A source of CRe, i.e. $Q(t)$. 
\end{itemize}
The time evolution of the particle spectrum as well as the resulting synchrotron emission have to be estimated numerically. \par
While semi-analytical models have been applied successfully in the past \citep[e.g.][]{2003ApJ...584..190F} the complexity of the problem really requires a fully numerical approach.

\subsection{First Simulations}
We show results from a first numerical study of reacceleration in cluster mergers \citep{2012arXiv1211.3337D}. 
This work makes use of the MHDSPH code {\small GADGET3} \citep{2005MNRAS.364.1105S,2009MNRAS.398.1678D} and a model for isolated cluster collisions (Donnert in prep., section \ref{sect.ICs}). We self-consistently follow the collisionless dynamics of dark matter (DM), and the gas in the MHD-approximation. This simulates the temperature, density and magnetic field distribution of the merger over time. In addition the following parameters are used:
\begin{itemize}
    \item the magnetic field is setup with $\eta_\mathrm{mag}=1$
    \item Turbulence follows a Kraichnan spectrum below the numerical resolution, so the diffusion coefficient equation \ref{eq.dpp} is used. Above the resolution scale turbulence is computed explicitely by the code.
    \item A numerical model for smoothed particle hydrodynamics (SPH) turbulence gives $v_\mathrm{turb}$ on the kernel scale $l = 2h_\mathrm{sml}$. Here $h_\mathrm{sml}$ is the compact support of the kernel.
    \item $\eta_\mathrm{turb} = 0.2$
    \item CRe are assumed to be constantly injected into the ICM. The injection coefficient is modelled as:
        \begin{align}
            Q_{\mathrm{e}}(\gamma,t) = K_e \varepsilon_\mathrm{th}^2\gamma^{-2} \sqrt{1 - \frac{\gamma}{10^5}}  (1 - \frac{50}{\gamma})\label{eq.numinj},
        \end{align}
        which is a power-law with smooth cut-offs at $\gamma = p/\mathrm{m}_\mathrm{e}c = 50,10^5 $. As the injection scales as the thermal energy density $\varepsilon_\mathrm{th}$ it is equivalent to an hadronic injection. $K_\mathrm{e} \approx 10^{-4}$ 
\end{itemize}

\subsubsection{Fokker-Planck Code}\label{sect.fkp}
\begin{figure}
	\centering
	\includegraphics[width=0.45\textwidth]{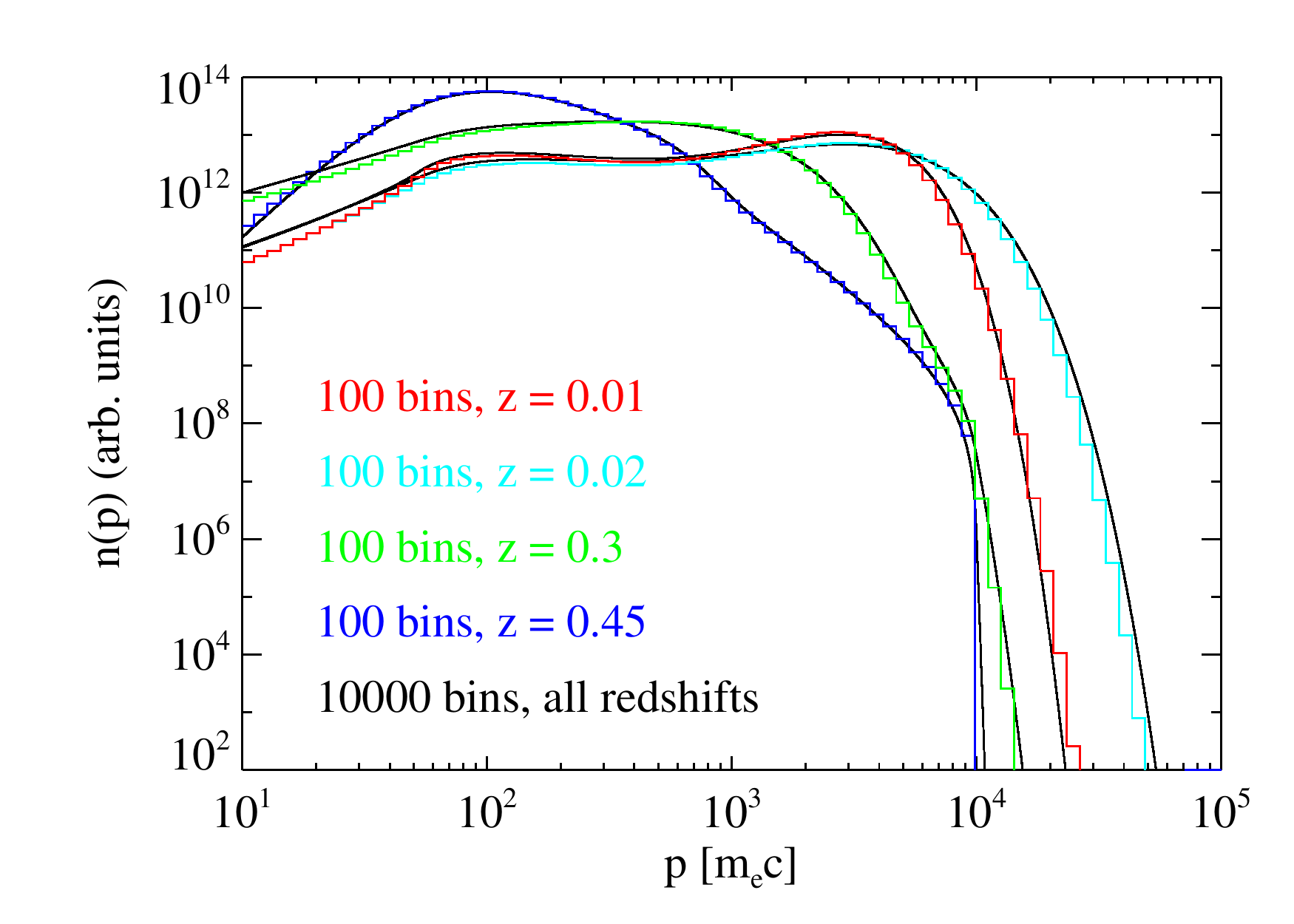}
	\caption{Comparison of CRe spectra from our code (colours) using 100 gridcells with another code (black) used in \citet{2005MNRAS.357.1313C} with 10000 cells.} \label{img.convergence}
\end{figure}

In post-processing to the simulation we follow dynamics of the CRe spectrum by solving the transport equation \ref{eqn.fkp} (Donnert in prep.) in parallel. We linearly interpolate between simulation outputs every 10 Myr and use a conservative timestep of 1 Myr in the code. The Fokker-Planck coefficients are computed from equations  \ref{eq.radlosses}, \ref{eq.ilosses}, \ref{eq.dpp} and \ref{eq.numinj} using the thermal and turbulence properties from the simulation. \par
We sample the CRe spectrum in 100 logarithmic bins between $p/m_\mathrm{e}c = 1$ and $10^6$. We realise open boundary conditions with the method from \citet{1986ApJ...308..929B}. We use the method of \citet{1970CompPhys.ChangCooper} to compute the time evolution of the spectrum. They derive a first-order accurate adaptive upwind scheme which can be analytically shown to:
\begin{itemize}
    \item be unconditionally stable
    \item guarantee positivity
    \item converge to the steady-state solution.
	\item conserve particle number.
\end{itemize}
This means it is ideally suited for our purpose to compute millions of spectra on particles of an SPH simulation. Specifically it allows us to use a logarithmic grid with only a small number of cells. The convergence and accuracy is tested against a run using the Fokker-Planck code from \citet{2005MNRAS.357.1313C}. In figure \ref{img.convergence} we compare results from their code using 10000 gridcells (black) with ours using 100 gridcells (colours). The input data are taken from \citet{2005MNRAS.357.1313C}. Deviations at low momenta occur due different implementations of the open boundary conditions. \par
\subsubsection{Thermal Model}\label{sect.ICs}
\begin{table}
    \centering
    \begin{tabular}{c c l l}
       Value & Unit & Cluster 0 & Cluster 1 \\\hline
       $a_\mathrm{Hernq}$ & kpc & 811  & 473 \\
       $r_{200}$ & kpc & 2192 & 1380 \\
       $r_\mathrm{core}$  & kpc & 237 & 130 \\
       $\mathrm{M}_{200}$ & $10^{15} \mathrm{M}_\odot $ & 1.33 & 0.17 \\
       $\rho_\mathrm{0,gas}$ &  $10^{-26}\mathrm{g}/ \mathrm{cm}^3$ & 0.9 & 1.1 \\
       $\mathrm{T}(r_\mathrm{core})$ & $10^7 \mathrm{K}$ & 9.2  &  2.8 
    \end{tabular}
    \caption{Model parameters of the initial conditions based on Donnert in prep. }\label{tab.ic}
\end{table}
We use an idealised model for major cluster merger based on the Hernquist profile \citep{1990ApJ...356..359H} for the collisionless matter. We identify a Hernquist profile with an NFW-profile \citep{1996ApJ...462..563N} with concentration parameter $c$ according to \citet{2005ApJ...620L..79S} The ICM is modelled as a $\beta$-model \citep{1966AJ.....71...64K,1978A&A....70..677C} with $\beta=2/3$ and $r_\mathrm{core} = c/3$. The hydrostatic equation can then be solved analytically to give a roughly constant temperature profile (Donnert in prep.). \par
We simulate a merging system with a mass ratio of 1:8, total mass of $1.5\times 10^{15}\,\mathrm{M}_\odot$ and a baryon fraction $b_f=0.17$. The two clusters are relaxed separately and then joined in a periodic box of 10 Mpc size. They are set on a zero energy orbit with an impact parameter of 300 Mpc. The magnetic field is set-up divergence free in k-space from a Gaussian random vector potential. This is transformed to a grid of 150 kpc size and the particles are initialised via NGP sampling. The field is attenuated according to $\eta_\mathrm{mag}=1$, which introduces divergence. We rely on the divergence cleaning of the code to deal with this in the beginning of the simulation.

\subsubsection{SPH-algorithm \& Turbulence}\label{sect.SPHturb}
To catch the bulk of turbulence generated in our simulation we use an SPH algorthim tuned towards minimising viscosity in the flow. We use a time-dependent viscosity approach \citep{2005MNRAS.364..753D} and a description for artificial thermal conduction and magnetic diffusion to resolve instabilities \citep{2007arXiv0709.2772P}. We use the high-order C4 Wendland kernel with 210 kernel-weighted neighbours \citep{2012MNRAS.425.1068D}. \par
The local turbulent energy is estimated from the RMS velocity dispersion of neighbours within the SPH kernel. By using a kernel with large compact support we make sure that viscous damping happens within the kernel scale. We set the low viscosity scheme to retain a minimum viscosity, corresponding to $\alpha = 0.01$ in the viscosity formulation of \citet{2005MNRAS.364..753D}. This way our approach remains conservative.

\subsubsection{Results}

\begin{figure*}
	\centering
	\includegraphics[width=\textwidth]{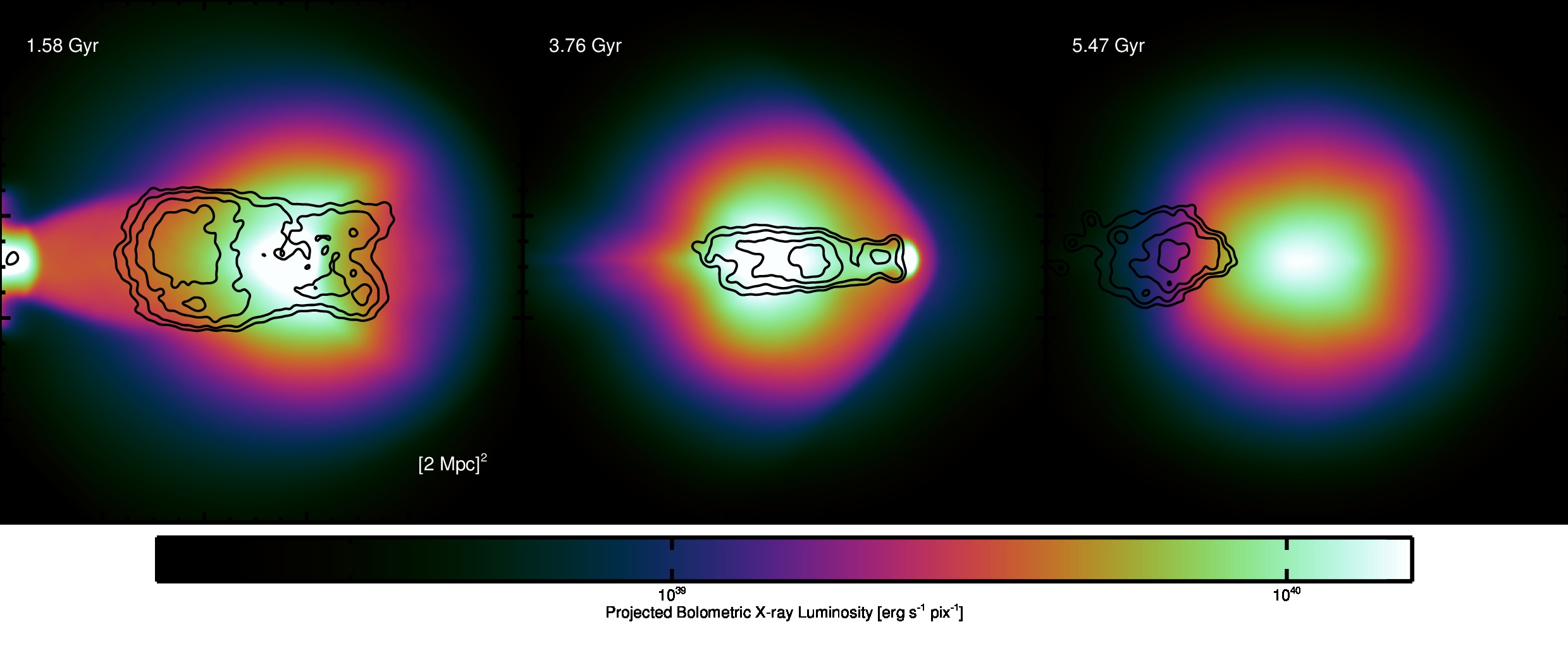}
	\caption{Bolometric X-ray emission from the system with radio synchrotron emission at 1.4 GHz overlayed as contours \citep{2012arXiv1211.3337D}. We show three different times: halo phase after the first (left) and second (middle) merger and off-center emission from an infalling stream (right). } \label{img.radioem}
\end{figure*}

We use our parallel projection routine {\small SMAC2} to extract thermal and non-thermal synthetic observations from the system. In figure \ref{img.radioem} we show projections of the X-ray emission at three different radio bright stages of the system. \par
\paragraph{The thermal evolution} of the system can be summarised as follows: 
\begin{itemize}
    \item Upon infall a large shock develops in front of the small cluster. This boosts the X-ray emission of the system, which peaks at the core passage (infall phase). 
    \item As the small DM core leaves the host cluster, the X-ray luminosity declines rapidly. The small core drags a part of the ICM along (effectively displacing the gas after the first encounter), causing turbulence (reacceleration phase).
    \item When the DM core reaches its turn-around point the turbulence has decayed, the ICM relaxed and the cluster is radio dark (decay phase).
\end{itemize}
These merging phases are repeated two times as the DM core oscillates in the host potential. The mass  of the sub-cluster successively declines and the host system is pre-disturbed. The core oscillations decay with time and the resulting emission peaks in the X-rays  decline.\par
Due to the impact parameter, the ICM of the host cluster recieves angular momentum and an infalling stream of hot gas develops after the second core passage. This has been seen in other simulations of this kind as well \citep{2001ApJ...561..621R}. The system relaxes after roughly 6 Gyr. \par
\paragraph{The non-thermal emission} follows the spatial and temporal evolution of turbulence in the simulation. Shortly after the first passage of the smaller DM core a large volume filling off-center halo develops (fig. \ref{img.radioem}, left), which decays within 1 Gyr. After the second passage an elongated smaller halo centered on the peak X-ray emission of the pre-disturbed cluster can be seen (fig. \ref{img.radioem}, middle). After relaxation of the DM cores, large off-center emission develops on top of the turbulent infall stream (fig. \ref{img.radioem}, right).  In between these phases the system becomes radio quiet as expected from analytical models. \par
\begin{figure*}
	\centering
	\includegraphics[width=0.32\textwidth]{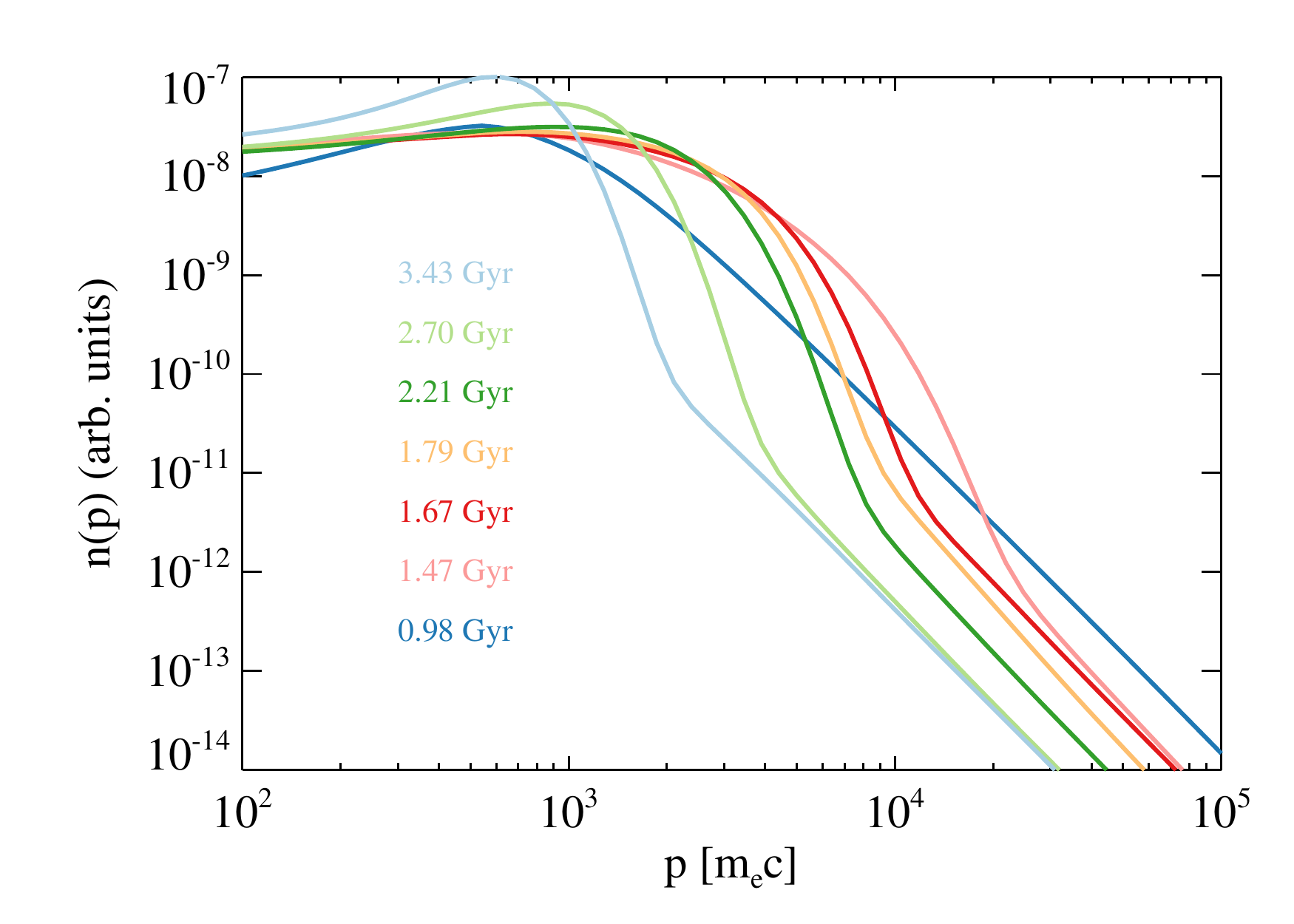}
	\includegraphics[width=0.32\textwidth]{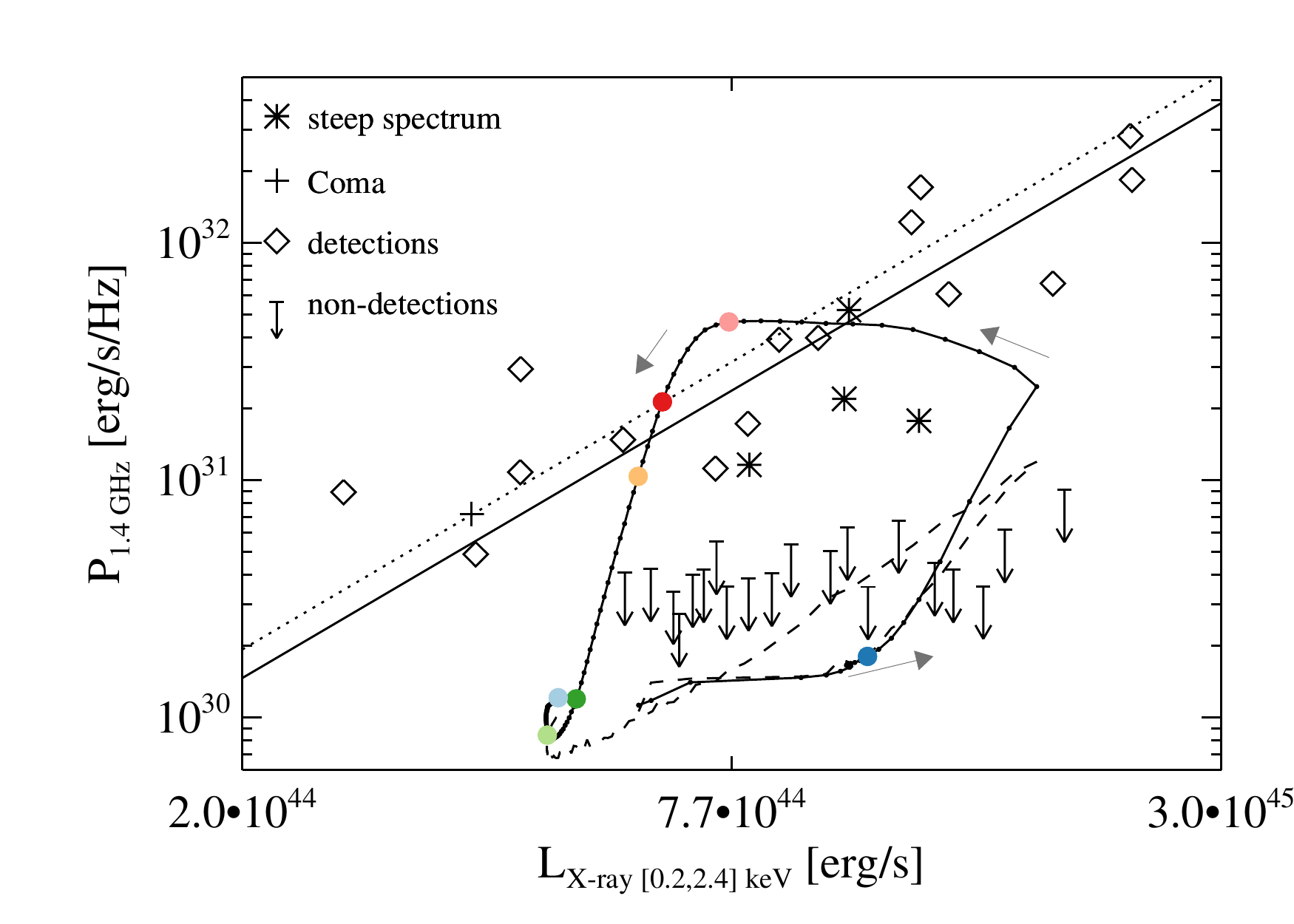}
	\includegraphics[width=0.32\textwidth]{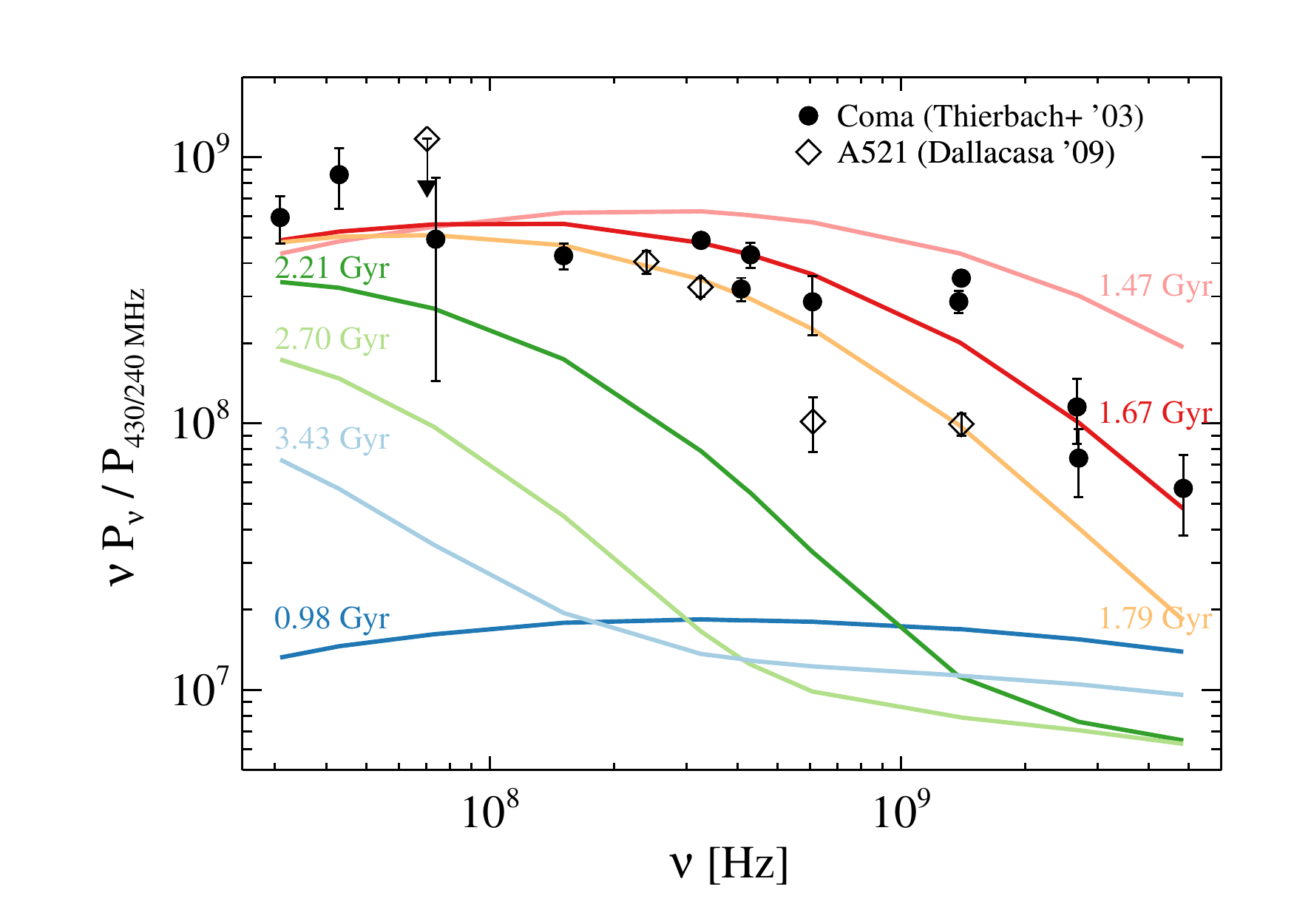}
	\caption{Left: Exemplary CRe spectrum of a particle in the simulation at different times. Middle: Observed radio-X-ray correlation (black line) and observed halos (diamonds, Coma: cross, ultra-steep: asterisks), upper limits (arrows). Simulated system (black curve, 10 Myr intervalls marked), simulated injection only (dashed curve). Colored dots correspond to the curves shown in the other panels. Right: Observed radio spectra of Coma (dots) and A521 (diamonds) and of the system at different times.} \label{img.reaccresults}
\end{figure*}

\paragraph{For a more quantative analysis} we will focus on the first passage only. In figure \ref{img.reaccresults} (left) we show CRe electron spectra from a single particle in the simulation at seven times between 1 Gyr and 3.5 Gyrs. At 1 Gyr the CRe spectrum is in the equillibrium state between injection and cooling (blue). At 1.5 Gyr (maximum halo brightness) the spectrum shows the typical bending from turbulent reacceleration (light red). In the following times the population cools, i.e. reacceleration is not efficient anymore and CRe diffuse to synchrotron dark momenta. \par
In figure \ref{img.reaccresults}, middle panel, we show the evolution of the system in the P14-LX plane (black line). We mark every 0.2 Myr with a small dot on the line and the times shown in the other two plots with corresponding colours. The injection only model is shown as a dashed line. The observed correlation (straight black line), observed halos (diamonds) and ultra-steep spectrum halos (asterisks) and upper-limits (arrows) are added as well. In this model, the system is below the upper-limits in its initial equillibrium state (blue dot). Upon infall the shock rapidly increases X-ray and synchrotron luminosity. In the reacceleration phase the X-ray luminosity declines, because the DM core drags ICM gas out of the host system. This causes turbulence and reacceleration, which boosts the radio luminosity to maximum brightness and on the correlation (light red point). As turbulence decays the radio emission declines and the cluster leaves the correlation. Within one Gyr the radio luminosity falls below the upper limits again. \par
In figure \ref{img.reaccresults}, right panel we show synchrotron spectra of the system. The colours correspond to the times discussed before. The synchrotron spectrum at 1Gyr (dark blue) is consistent with the analytical expectation for the equillibrium of cooling and injection (eq. \ref{eq.hadsol}). At its maximum brightness of the system, its radio spectrum is flatter than the observed Coma spectrum. The simulated spectrum then steepens with time during the cooling phase. At 1.7 Gyr it fits the Coma spectrum (filled dots) and at 1.8 Gyr the ultra-steep spectrum of A521 (diamonds). With time the spectrum flattens again and approaches the equillibrium spectrum at 3.4 Gyrs.

\section{Conclusions}
We reported on the status of models for giant radio halos. Reviewing the current picture on radio halos from observations, we presented a short introduction to hadronic models. Using a pure hadronic model in a cosmological MHD simulation, we argued that hadronic models fail to \emph{simultaneously} reproduce key observables:
\begin{itemize}
    \item the Coma radio profile within the non-thermal pressure constrains,
    \item the Coma Compton-y - radio correlation
    \item the Coma spectrum and radio brightness within the current $\gamma$-ray upper limits,
    \item the bimodal distribution in cluster brightness.
\end{itemize}
We put forward three theoretical arguments against non-classical hadronic (streaming) models:
\begin{itemize}
    \item Galaxy clusters are unlikely to be free of turbulence on small scales due to high Reynolds numbers and constant infall of small DM haloes.
    \item Efficient damping does not imply complete absense of turbulence if a cascade is present.
    \item Linear theory is not applicable for non-negligable damping. In turn non-linear theory predicts efficient CR scattering also for small pitch angles.
\end{itemize}
Considering this evidence we presented as short introduction to reacceleration models. We motivated the transport equation and its terms governing cooling, injection and reacceleration of the CRe population. We presented first results from a numerical approach to the problem in the framework of idealised cluster collisions. Using a constant injection of CRe, the idealised system reproduces key observables, specifically:
\begin{itemize}
    \item the variety of observed spectral indices, from flat to ultra-steep
    \item the transient nature of radio halos
\end{itemize}
This reaffirms prior expectations from theoretical approaches. Naturally this idealised model is only a first step towards a more detailed modelling of the non-thermal components of the ICM. Upcoming radio telescopes like LOFAR will test our current understanding of CR dynamics and detailed numerical modelling in a cosmological framework seems highly appropriate. \par
We would like to emphasize that future approaches must include CR protons as well, and the absense of any clear observational signature of these particles remains an exciting problem in the next decades. One may also consider that state-of-the-art models for  radio haloes require a  plasma, \emph{not} a fluid model of the ICM on small scales. This comes with significant complications in the physics, but bares the chance of a more complete understanding of the microphysics of the ICM, a truly unique plasma.

\acknowledgements
The author thanks the German Astronomical Society for the PhD award 2012. This work has been done with the support of many people, special thanks go to K.Dolag and G.Brunetti. J.D. acknowledges support by PRIN-INAF2009 and the FP7 Marie Curie programme 'People' of the European Union. The computations where performed at the ``Rechenzentrum der Max-Planck-Gesellschaft'', with resources assigned to the ``Max-Planck-Institut f\"ur Astrophysik''.

\bibliographystyle{mn2e} \bibliography{master}

\begin{thebibliography}{}

\bibitem[\protect\citeauthoryear{{Achterberg}}{{Achterberg}}{1981}]{1981A&A...%
.98..161A}
{Achterberg} A.,  1981, \aap, 98, 161

\bibitem[\protect\citeauthoryear{{Ackermann}}{{Ackermann}}{2010}]{2010ApJ...71%
7L..71A}
{Ackermann} M. e.~a.,  2010, \apjl, 717, L71

\bibitem[\protect\citeauthoryear{{Aharonian}}{{Aharonian}}{2009}]{2009A&A...50%
2..437A}
{Aharonian} F. e.~a.,  2009, \aap, 502, 437

\bibitem[\protect\citeauthoryear{{Basu}}{{Basu}}{2012}]{2012MNRAS.421L.112B}
{Basu} K.,  2012, \mnras, 421, L112

\bibitem[\protect\citeauthoryear{{Beck}, {Lesch}, {Dolag}, {Kotarba}, {Geng} \&
  {Stasyszyn}}{{Beck} et~al.}{2012}]{2012MNRAS.422.2152B}
{Beck} A.~M.,  {Lesch} H.,  {Dolag} K.,  {Kotarba} H.,  {Geng} A.,
  {Stasyszyn} F.~A.,  2012, \mnras, 422, 2152

\bibitem[\protect\citeauthoryear{{Ben-Israel}, {Piran}, {Eviatar} \&
  {Weinstock}}{{Ben-Israel} et~al.}{1975}]{1975Ap&SS..38..125B}
{Ben-Israel} I.,  {Piran} T.,  {Eviatar} A.,    {Weinstock} J.,  1975, \apss,
  38, 125

\bibitem[\protect\citeauthoryear{{Berezinsky}, {Blasi} \&
  {Ptuskin}}{{Berezinsky} et~al.}{1997}]{1997ApJ...487..529B}
{Berezinsky} V.~S.,  {Blasi} P.,    {Ptuskin} V.~S.,  1997, \apj, 487, 529

\bibitem[\protect\citeauthoryear{{Blasi}}{{Blasi}}{2010}]{2010MNRAS.402.2807B}
{Blasi} P.,  2010, \mnras, 402, 2807

\bibitem[\protect\citeauthoryear{{Blasi}, {Amato} \& {Caprioli}}{{Blasi}
  et~al.}{2007}]{2007MNRAS.375.1471B}
{Blasi} P.,  {Amato} E.,    {Caprioli} D.,  2007, \mnras, 375, 1471

\bibitem[\protect\citeauthoryear{{Blasi} \& {Colafrancesco}}{{Blasi} \&
  {Colafrancesco}}{1999}]{1999APh....12..169B}
{Blasi} P.,  {Colafrancesco} S.,  1999, Astroparticle Physics, 12, 169

\bibitem[\protect\citeauthoryear{{Blasi}, {Gabici} \& {Brunetti}}{{Blasi}
  et~al.}{2007a}]{2007IJMPA..22..681B}
{Blasi} P.,  {Gabici} S.,    {Brunetti} G.,  2007a, International Journal of
  Modern Physics A, 22, 681

\bibitem[\protect\citeauthoryear{{Blasi}, {Gabici} \& {Brunetti}}{{Blasi}
  et~al.}{2007b}]{2007astro.ph..1545B}
{Blasi} P.,  {Gabici} S.,    {Brunetti} G.,  2007b, ArXiv Astrophysics e-prints

\bibitem[\protect\citeauthoryear{{Bonafede}, {Feretti}, {Murgia}, {Govoni},
  {Giovannini}, {Dallacasa}, {Dolag} \& {Taylor}}{{Bonafede}
  et~al.}{2010}]{2010A&A...513A..30B}
{Bonafede} A.,  {Feretti} L.,  {Murgia} M.,  {Govoni} F.,  {Giovannini} G.,
  {Dallacasa} D.,  {Dolag} K.,    {Taylor} G.~B.,  2010, \aap, 513, A30+

\bibitem[\protect\citeauthoryear{{Borovsky} \& {Eilek}}{{Borovsky} \&
  {Eilek}}{1986}]{1986ApJ...308..929B}
{Borovsky} J.~E.,  {Eilek} J.~A.,  1986, \apj, 308, 929

\bibitem[\protect\citeauthoryear{{Brown}, {Emerick}, {Rudnick} \&
  {Brunetti}}{{Brown} et~al.}{2011}]{2011ApJ...740L..28B}
{Brown} S.,  {Emerick} A.,  {Rudnick} L.,    {Brunetti} G.,  2011, \apjl, 740,
  L28

\bibitem[\protect\citeauthoryear{{Brown} \& {Rudnick}}{{Brown} \&
  {Rudnick}}{2011}]{2011MNRAS.412....2B}
{Brown} S.,  {Rudnick} L.,  2011, \mnras, 412, 2

\bibitem[\protect\citeauthoryear{{Br{\"u}ggen}, {van Weeren} \&
  {R{\"o}ttgering}}{{Br{\"u}ggen} et~al.}{2012}]{2012MNRAS.425L..76B}
{Br{\"u}ggen} M.,  {van Weeren} R.~J.,    {R{\"o}ttgering} H.~J.~A.,  2012,
  \mnras, 425, L76

\bibitem[\protect\citeauthoryear{{Brunetti} \& {Blasi}}{{Brunetti} \&
  {Blasi}}{2005}]{2005MNRAS.363.1173B}
{Brunetti} G.,  {Blasi} P.,  2005, \mnras, 363, 1173

\bibitem[\protect\citeauthoryear{{Brunetti}, {Blasi}, {Cassano} \&
  {Gabici}}{{Brunetti} et~al.}{2004}]{2004MNRAS.350.1174B}
{Brunetti} G.,  {Blasi} P.,  {Cassano} R.,    {Gabici} S.,  2004, \mnras, 350,
  1174

\bibitem[\protect\citeauthoryear{{Brunetti}, {Blasi}, {Reimer}, {Rudnick},
  {Bonafede} \& {Brown}}{{Brunetti} et~al.}{2012}]{2012MNRAS.426..956B}
{Brunetti} G.,  {Blasi} P.,  {Reimer} O.,  {Rudnick} L.,  {Bonafede} A.,
  {Brown} S.,  2012, \mnras, 426, 956

\bibitem[\protect\citeauthoryear{{Brunetti}, {Cassano}, {Dolag} \&
  {Setti}}{{Brunetti} et~al.}{2009}]{2009A&A...507..661B}
{Brunetti} G.,  {Cassano} R.,  {Dolag} K.,    {Setti} G.,  2009, \aap, 507, 661

\bibitem[\protect\citeauthoryear{{Brunetti}, {Giacintucci}, {Cassano}, {Lane},
  {Dallacasa}, {Venturi}, {Kassim}, {Setti}, {Cotton} \&
  {Markevitch}}{{Brunetti} et~al.}{2008}]{2008Natur.455..944B}
{Brunetti} G.,  {Giacintucci} S.,  {Cassano} R.,  {Lane} W.,  {Dallacasa} D.,
  {Venturi} T.,  {Kassim} N.~E.,  {Setti} G.,  {Cotton} W.~D.,    {Markevitch}
  M.,  2008, \nat, 455, 944

\bibitem[\protect\citeauthoryear{{Brunetti} \& {Lazarian}}{{Brunetti} \&
  {Lazarian}}{2007}]{2007MNRAS.378..245B}
{Brunetti} G.,  {Lazarian} A.,  2007, \mnras, 378, 245

\bibitem[\protect\citeauthoryear{{Brunetti} \& {Lazarian}}{{Brunetti} \&
  {Lazarian}}{2011a}]{2011MNRAS.410..127B}
{Brunetti} G.,  {Lazarian} A.,  2011a, \mnras, 410, 127

\bibitem[\protect\citeauthoryear{{Brunetti} \& {Lazarian}}{{Brunetti} \&
  {Lazarian}}{2011b}]{2011MNRAS.412..817B}
{Brunetti} G.,  {Lazarian} A.,  2011b, \mnras, 412, 817

\bibitem[\protect\citeauthoryear{{Brunetti}, {Setti}, {Feretti} \&
  {Giovannini}}{{Brunetti} et~al.}{2001}]{2001MNRAS.320..365B}
{Brunetti} G.,  {Setti} G.,  {Feretti} L.,    {Giovannini} G.,  2001, \mnras,
  320, 365

\bibitem[\protect\citeauthoryear{{Cassano} \& {Brunetti}}{{Cassano} \&
  {Brunetti}}{2005}]{2005MNRAS.357.1313C}
{Cassano} R.,  {Brunetti} G.,  2005, \mnras, 357, 1313

\bibitem[\protect\citeauthoryear{{Cassano}, {Brunetti}, {Norris},
  {Roettgering}, {Johnston-Hollitt} \& {Trasatti}}{{Cassano}
  et~al.}{2012}]{2012arXiv1210.1020C}
{Cassano} R.,  {Brunetti} G.,  {Norris} R.~P.,  {Roettgering} H.~J.~A.,
  {Johnston-Hollitt} M.,    {Trasatti} M.,  2012, ArXiv e-prints

\bibitem[\protect\citeauthoryear{{Cassano}, {Brunetti}, {R{\"o}ttgering} \&
  {Br{\"u}ggen}}{{Cassano} et~al.}{2010}]{2010A&A...509A..68C}
{Cassano} R.,  {Brunetti} G.,  {R{\"o}ttgering} H.~J.~A.,    {Br{\"u}ggen} M.,
  2010, \aap, 509, A68+

\bibitem[\protect\citeauthoryear{{Cassano}, {Brunetti}, {Setti}, {Govoni} \&
  {Dolag}}{{Cassano} et~al.}{2007}]{2007MNRAS.378.1565C}
{Cassano} R.,  {Brunetti} G.,  {Setti} G.,  {Govoni} F.,    {Dolag} K.,  2007,
  \mnras, 378, 1565

\bibitem[\protect\citeauthoryear{{Cassano}, {Ettori}, {Giacintucci},
  {Brunetti}, {Markevitch}, {Venturi} \& {Gitti}}{{Cassano}
  et~al.}{2010}]{2010ApJ...721L..82C}
{Cassano} R.,  {Ettori} S.,  {Giacintucci} S.,  {Brunetti} G.,  {Markevitch}
  M.,  {Venturi} T.,    {Gitti} M.,  2010, \apjl, 721, L82

\bibitem[\protect\citeauthoryear{{Cavaliere} \& {Fusco-Femiano}}{{Cavaliere} \&
  {Fusco-Femiano}}{1978}]{1978A&A....70..677C}
{Cavaliere} A.,  {Fusco-Femiano} R.,  1978, \aap, 70, 677

\bibitem[\protect\citeauthoryear{{Cavaliere}, {Gursky} \& {Tucker}}{{Cavaliere}
  et~al.}{1971}]{1971Natur.231..437C}
{Cavaliere} A.~G.,  {Gursky} H.,    {Tucker} W.~H.,  1971, \nat, 231, 437

\bibitem[\protect\citeauthoryear{{Chang} \& {Cooper}}{{Chang} \&
  {Cooper}}{1970}]{1970CompPhys.ChangCooper}
{Chang} J.,  {Cooper} G.,  1970, Journal of Computational Physics, 6, 1

\bibitem[\protect\citeauthoryear{{Churazov}, {Vikhlinin}, {Zhuravleva},
  {Schekochihin}, {Parrish}, {Sunyaev}, {Forman}, {B{\"o}hringer} \&
  {Randall}}{{Churazov} et~al.}{2012}]{2012MNRAS.421.1123C}
{Churazov} E.,  {Vikhlinin} A.,  {Zhuravleva} I.,  {Schekochihin} A.,
  {Parrish} I.,  {Sunyaev} R.,  {Forman} W.,  {B{\"o}hringer} H.,    {Randall}
  S.,  2012, \mnras, 421, 1123

\bibitem[\protect\citeauthoryear{{Dallacasa}, {Brunetti}, {Giacintucci},
  {Cassano}, {Venturi}, {Macario}, {Kassim}, {Lane} \& {Setti}}{{Dallacasa}
  et~al.}{2009}]{2009ApJ...699.1288D}
{Dallacasa} D.,  {Brunetti} G.,  {Giacintucci} S.,  {Cassano} R.,  {Venturi}
  T.,  {Macario} G.,  {Kassim} N.~E.,  {Lane} W.,    {Setti} G.,  2009, \apj,
  699, 1288

\bibitem[\protect\citeauthoryear{{Dehnen} \& {Aly}}{{Dehnen} \&
  {Aly}}{2012}]{2012MNRAS.425.1068D}
{Dehnen} W.,  {Aly} H.,  2012, \mnras, 425, 1068

\bibitem[\protect\citeauthoryear{{Deiss}, {Reich}, {Lesch} \&
  {Wielebinski}}{{Deiss} et~al.}{1997}]{1997A&A...321...55D}
{Deiss} B.~M.,  {Reich} W.,  {Lesch} H.,    {Wielebinski} R.,  1997, \aap, 321,
  55

\bibitem[\protect\citeauthoryear{{Dennison}}{{Dennison}}{1980}]{1980ApJ...239L%
..93D}
{Dennison} B.,  1980, \apjl, 239, L93

\bibitem[\protect\citeauthoryear{{Dermer}}{{Dermer}}{1986}]{1986A&A...157..223%
D}
{Dermer} C.~D.,  1986, \aap, 157, 223

\bibitem[\protect\citeauthoryear{{Dolag}, {Bartelmann} \& {Lesch}}{{Dolag}
  et~al.}{1999}]{1999A&A...348..351D}
{Dolag} K.,  {Bartelmann} M.,    {Lesch} H.,  1999, \aap, 348, 351

\bibitem[\protect\citeauthoryear{{Dolag} \& {Ensslin}}{{Dolag} \&
  {Ensslin}}{2000}]{2000A&A...362..151D}
{Dolag} K.,  {Ensslin} T.~A.,  2000, \aap, 362, 151

\bibitem[\protect\citeauthoryear{{Dolag}, {Schindler}, {Govoni} \&
  {Feretti}}{{Dolag} et~al.}{2001}]{2001A&A...378..777D}
{Dolag} K.,  {Schindler} S.,  {Govoni} F.,    {Feretti} L.,  2001, \aap, 378,
  777

\bibitem[\protect\citeauthoryear{{Dolag} \& {Stasyszyn}}{{Dolag} \&
  {Stasyszyn}}{2009}]{2009MNRAS.398.1678D}
{Dolag} K.,  {Stasyszyn} F.,  2009, \mnras, 398, 1678

\bibitem[\protect\citeauthoryear{{Dolag}, {Vazza}, {Brunetti} \&
  {Tormen}}{{Dolag} et~al.}{2005}]{2005MNRAS.364..753D}
{Dolag} K.,  {Vazza} F.,  {Brunetti} G.,    {Tormen} G.,  2005, \mnras, 364,
  753

\bibitem[\protect\citeauthoryear{{Donnert}, {Dolag}, {Bonafede}, {Cassano} \&
  {Brunetti}}{{Donnert} et~al.}{2010}]{2010MNRAS.401...47D}
{Donnert} J.,  {Dolag} K.,  {Bonafede} A.,  {Cassano} R.,    {Brunetti} G.,
  2010, \mnras, 401, 47

\bibitem[\protect\citeauthoryear{{Donnert}, {Dolag}, {Brunetti} \&
  {Cassano}}{{Donnert} et~al.}{2012}]{2012arXiv1211.3337D}
{Donnert} J.,  {Dolag} K.,  {Brunetti} G.,    {Cassano} R.,  2012, ArXiv
  e-prints

\bibitem[\protect\citeauthoryear{{Donnert}, {Dolag}, {Cassano} \&
  {Brunetti}}{{Donnert} et~al.}{2010}]{2010MNRAS.407.1565D}
{Donnert} J.,  {Dolag} K.,  {Cassano} R.,    {Brunetti} G.,  2010, \mnras, 407,
  1565

\bibitem[\protect\citeauthoryear{{Donnert}, {Dolag}, {Lesch} \&
  {M{\"u}ller}}{{Donnert} et~al.}{2009}]{2009MNRAS.392.1008D}
{Donnert} J.,  {Dolag} K.,  {Lesch} H.,    {M{\"u}ller} E.,  2009, \mnras, 392,
  1008

\bibitem[\protect\citeauthoryear{{Dupree}}{{Dupree}}{1966}]{1966PhFl....9.1773%
D}
{Dupree} T.~H.,  1966, Physics of Fluids, 9, 1773

\bibitem[\protect\citeauthoryear{{Eilek}}{{Eilek}}{1979}]{1979ApJ...230..373E}
{Eilek} J.~A.,  1979, \apj, 230, 373

\bibitem[\protect\citeauthoryear{{En{\ss}lin}, {Pfrommer}, {Miniati} \&
  {Subramanian}}{{En{\ss}lin} et~al.}{2011}]{2011A&A...527A..99E}
{En{\ss}lin} T.,  {Pfrommer} C.,  {Miniati} F.,    {Subramanian} K.,  2011,
  \aap, 527, A99+

\bibitem[\protect\citeauthoryear{{Ensslin}}{{Ensslin}}{2002}]{2002A&A...396L..%
17E}
{Ensslin} T.~A.,  2002, \aap, 396, L17

\bibitem[\protect\citeauthoryear{{Ensslin}, {Pfrommer}, {Springel} \&
  {Jubelgas}}{{Ensslin} et~al.}{2007}]{2007A&A...473...41E}
{Ensslin} T.~A.,  {Pfrommer} C.,  {Springel} V.,    {Jubelgas} M.,  2007, \aap,
  473, 41

\bibitem[\protect\citeauthoryear{{Eriksen}, {Hughes}, {Badenes}, {Fesen},
  {Ghavamian}, {Moffett}, {Plucinksy}, {Rakowski}, {Reynoso} \&
  {Slane}}{{Eriksen} et~al.}{2011}]{2011ApJ...728L..28E}
{Eriksen} K.~A.,  {Hughes} J.~P.,  {Badenes} C.,  {Fesen} R.,  {Ghavamian} P.,
  {Moffett} D.,  {Plucinksy} P.~P.,  {Rakowski} C.~E.,  {Reynoso} E.~M.,
  {Slane} P.,  2011, \apjl, 728, L28

\bibitem[\protect\citeauthoryear{{Felice} \& {Kulsrud}}{{Felice} \&
  {Kulsrud}}{2001}]{2001ApJ...553..198F}
{Felice} G.~M.,  {Kulsrud} R.~M.,  2001, \apj, 553, 198

\bibitem[\protect\citeauthoryear{{Feretti} \& {Giovannini}}{{Feretti} \&
  {Giovannini}}{1996}]{1996IAUS..175..333F}
{Feretti} L.,  {Giovannini} G.,  1996, in {R.~D.~Ekers, C.~Fanti, \&
  L.~Padrielli} ed., Extragalactic Radio Sources Vol.~175 of IAU Symposium,
  {Diffuse Cluster Radio Sources (Review)}.
pp 333--+

\bibitem[\protect\citeauthoryear{{Feretti}, {Giovannini}, {Govoni} \&
  {Murgia}}{{Feretti} et~al.}{2012}]{2012A&ARv..20...54F}
{Feretti} L.,  {Giovannini} G.,  {Govoni} F.,    {Murgia} M.,  2012, \aapr, 20,
  54

\bibitem[\protect\citeauthoryear{{Fermi}}{{Fermi}}{1949}]{1949PhRv...75.1169F}
{Fermi} E.,  1949, Physical Review, 75, 1169

\bibitem[\protect\citeauthoryear{{Fujita}, {Takizawa} \& {Sarazin}}{{Fujita}
  et~al.}{2003}]{2003ApJ...584..190F}
{Fujita} Y.,  {Takizawa} M.,    {Sarazin} C.~L.,  2003, \apj, 584, 190

\bibitem[\protect\citeauthoryear{{Gargat{\'e}} \& {Spitkovsky}}{{Gargat{\'e}}
  \& {Spitkovsky}}{2012}]{2012ApJ...744...67G}
{Gargat{\'e}} L.,  {Spitkovsky} A.,  2012, \apj, 744, 67

\bibitem[\protect\citeauthoryear{{Ginzburg} \& {Syrovatskii}}{{Ginzburg} \&
  {Syrovatskii}}{1965}]{1965ARA&A...3..297G}
{Ginzburg} V.~L.,  {Syrovatskii} S.~I.,  1965, \araa, 3, 297

\bibitem[\protect\citeauthoryear{{Giovannini}, {Bonafede}, {Feretti}, {Govoni},
  {Murgia}, {Ferrari} \& {Monti}}{{Giovannini}
  et~al.}{2009}]{2009A&A...507.1257G}
{Giovannini} G.,  {Bonafede} A.,  {Feretti} L.,  {Govoni} F.,  {Murgia} M.,
  {Ferrari} F.,    {Monti} G.,  2009, \aap, 507, 1257

\bibitem[\protect\citeauthoryear{{Goldstein}}{{Goldstein}}{1976}]{1976ApJ...20%
4..900G}
{Goldstein} M.~L.,  1976, \apj, 204, 900

\bibitem[\protect\citeauthoryear{{Govoni}, {Feretti}, {Giovannini},
  {B{\"o}hringer}, {Reiprich} \& {Murgia}}{{Govoni}
  et~al.}{2001}]{2001A&A...376..803G}
{Govoni} F.,  {Feretti} L.,  {Giovannini} G.,  {B{\"o}hringer} H.,  {Reiprich}
  T.~H.,    {Murgia} M.,  2001, \aap, 376, 803

\bibitem[\protect\citeauthoryear{{Hernquist}}{{Hernquist}}{1990}]{1990ApJ...35%
6..359H}
{Hernquist} L.,  1990, 356, 359

\bibitem[\protect\citeauthoryear{{Hoeft}, {Br{\"u}ggen} \& {Yepes}}{{Hoeft}
  et~al.}{2004}]{2004MNRAS.347..389H}
{Hoeft} M.,  {Br{\"u}ggen} M.,    {Yepes} G.,  2004, \mnras, 347, 389

\bibitem[\protect\citeauthoryear{{Hoeft}, {Br{\"u}ggen}, {Yepes},
  {Gottl{\"o}ber} \& {Schwope}}{{Hoeft} et~al.}{2008}]{2008MNRAS.391.1511H}
{Hoeft} M.,  {Br{\"u}ggen} M.,  {Yepes} G.,  {Gottl{\"o}ber} S.,    {Schwope}
  A.,  2008, \mnras, 391, 1511

\bibitem[\protect\citeauthoryear{{Holman}, {Ionson} \& {Scott}}{{Holman}
  et~al.}{1979}]{1979ApJ...228..576H}
{Holman} G.~D.,  {Ionson} J.~A.,    {Scott} J.~S.,  1979, \apj, 228, 576

\bibitem[\protect\citeauthoryear{{Iapichino} \& {Niemeyer}}{{Iapichino} \&
  {Niemeyer}}{2008}]{2008MNRAS.388.1089I}
{Iapichino} L.,  {Niemeyer} J.~C.,  2008, \mnras, 388, 1089

\bibitem[\protect\citeauthoryear{{Iapichino}, {Schmidt}, {Niemeyer} \&
  {Merklein}}{{Iapichino} et~al.}{2011}]{2011MNRAS.tmp..483I}
{Iapichino} L.,  {Schmidt} W.,  {Niemeyer} J.~C.,    {Merklein} J.,  2011,
  \mnras, pp 483--+

\bibitem[\protect\citeauthoryear{{Jaffe}}{{Jaffe}}{1977}]{1977ApJ...212....1J}
{Jaffe} W.~J.,  1977, \apj, 212, 1

\bibitem[\protect\citeauthoryear{{Jones}, {Birmingham} \& {Kaiser}}{{Jones}
  et~al.}{1978}]{1978PhFl...21..347J}
{Jones} F.~C.,  {Birmingham} T.~J.,    {Kaiser} T.~B.,  1978, Physics of
  Fluids, 21, 347

\bibitem[\protect\citeauthoryear{{Keshet} \& {Loeb}}{{Keshet} \&
  {Loeb}}{2010}]{2010ApJ...722..737K}
{Keshet} U.,  {Loeb} A.,  2010, \apj, 722, 737

\bibitem[\protect\citeauthoryear{{King}}{{King}}{1966}]{1966AJ.....71...64K}
{King} I.~R.,  1966, \aj, 71, 64

\bibitem[\protect\citeauthoryear{{Kuchar} \& {En{\ss}lin}}{{Kuchar} \&
  {En{\ss}lin}}{2011}]{2011A&A...529A..13K}
{Kuchar} P.,  {En{\ss}lin} T.~A.,  2011, \aap, 529, A13+

\bibitem[\protect\citeauthoryear{{Lazarian} \& {Brunetti}}{{Lazarian} \&
  {Brunetti}}{2011}]{2011MmSAI..82..636L}
{Lazarian} A.,  {Brunetti} G.,  2011, \memsai, 82, 636

\bibitem[\protect\citeauthoryear{{Longair}}{{Longair}}{1994}]{1994hea2.book...%
..L}
{Longair} M.~S.,  1994, {High energy astrophysics. Volume 2. Stars, the Galaxy
  and the interstellar medium.}

\bibitem[\protect\citeauthoryear{{Maier}, {Iapichino}, {Schmidt} \&
  {Niemeyer}}{{Maier} et~al.}{2009}]{2009ApJ...707...40M}
{Maier} A.,  {Iapichino} L.,  {Schmidt} W.,    {Niemeyer} J.~C.,  2009, \apj,
  707, 40

\bibitem[\protect\citeauthoryear{{Markevitch} \& {Vikhlinin}}{{Markevitch} \&
  {Vikhlinin}}{2007}]{2007PhR...443....1M}
{Markevitch} M.,  {Vikhlinin} A.,  2007, \physrep, 443, 1

\bibitem[\protect\citeauthoryear{{Mathis}, {Lemson}, {Springel}, {Kauffmann},
  {White}, {Eldar} \& {Dekel}}{{Mathis} et~al.}{2002}]{2002MNRAS.333..739M}
{Mathis} H.,  {Lemson} G.,  {Springel} V.,  {Kauffmann} G.,  {White} S.~D.~M.,
  {Eldar} A.,    {Dekel} A.,  2002, \mnras, 333, 739

\bibitem[\protect\citeauthoryear{{Mazzotta} \& {Planck
  Collaboration}}{{Mazzotta} \& {Planck
  Collaboration}}{2012}]{2012AAS...22050705M}
{Mazzotta} P.,  {Planck Collaboration} 2012, in American Astronomical Society
  Meeting Abstracts \#220 Vol.~220 of American Astronomical Society Meeting
  Abstracts, {Planck Intermediate Paper: Physics Of The Hot Gas In The Coma
  Cluster}.
p. 507.05

\bibitem[\protect\citeauthoryear{{Meekins}, {Fritz}, {Chubb} \&
  {Friedman}}{{Meekins} et~al.}{1971}]{1971Natur.231..107M}
{Meekins} J.~F.,  {Fritz} G.,  {Chubb} T.~A.,    {Friedman} H.,  1971, \nat,
  231, 107

\bibitem[\protect\citeauthoryear{{Melrose}}{{Melrose}}{1968}]{1968Ap&SS...2..1%
71M}
{Melrose} D.~B.,  1968, \apss, 2, 171

\bibitem[\protect\citeauthoryear{{Melrose}}{{Melrose}}{1980}]{1980panp.book...%
..M}
{Melrose} D.~B.,  1980, {Plasma astrohysics. Nonthermal processes in diffuse
  magnetized plasmas - Vol.1: The emission, absorption and transfer of waves in
  plasmas; Vol.2: Astrophysical applications}

\bibitem[\protect\citeauthoryear{{Miniati}, {Jones}, {Kang} \& {Ryu}}{{Miniati}
  et~al.}{2001}]{2001ApJ...562..233M}
{Miniati} F.,  {Jones} T.~W.,  {Kang} H.,    {Ryu} D.,  2001, \apj, 562, 233

\bibitem[\protect\citeauthoryear{{Navarro}, {Frenk} \& {White}}{{Navarro}
  et~al.}{1996}]{1996ApJ...462..563N}
{Navarro} J.~F.,  {Frenk} C.~S.,    {White} S.~D.~M.,  1996, \apj, 462, 563

\bibitem[\protect\citeauthoryear{{Pacholczyk}}{{Pacholczyk}}{1970}]{1970ranp.b%
ook.....P}
{Pacholczyk} A.~G.,  1970, {Radio astrophysics. Nonthermal processes in
  galactic and extragalactic sources}

\bibitem[\protect\citeauthoryear{{Perkins}}{{Perkins}}{2008}]{2008AIPC.1085..5%
69P}
{Perkins} J.~S.,  2008, in American Institute of Physics Conference Series
  Vol.~1085 of American Institute of Physics Conference Series, {VERITAS
  Observations of the Coma Cluster of Galaxies}.
pp 569--572

\bibitem[\protect\citeauthoryear{{Petrosian}}{{Petrosian}}{2001}]{2001ApJ...55%
7..560P}
{Petrosian} V.,  2001, \apj, 557, 560

\bibitem[\protect\citeauthoryear{{Pfrommer} \& {Dursi}}{{Pfrommer} \&
  {Dursi}}{2010}]{2010NatPh...6..520P}
{Pfrommer} C.,  {Dursi} J.,  2010, Nature Physics, 6, 520

\bibitem[\protect\citeauthoryear{{Pfrommer} \& {Ensslin}}{{Pfrommer} \&
  {Ensslin}}{2004}]{2004A&A...413...17P}
{Pfrommer} C.,  {Ensslin} T.~A.,  2004, \aap, 413, 17

\bibitem[\protect\citeauthoryear{{Pfrommer} \& {En{\ss}lin}}{{Pfrommer} \&
  {En{\ss}lin}}{2004}]{2004MNRAS.352...76P}
{Pfrommer} C.,  {En{\ss}lin} T.~A.,  2004, \mnras, 352, 76

\bibitem[\protect\citeauthoryear{{Pfrommer}, {Ensslin} \&
  {Springel}}{{Pfrommer} et~al.}{2008}]{2008MNRAS.385.1211P}
{Pfrommer} C.,  {Ensslin} T.~A.,    {Springel} V.,  2008, \mnras, 385, 1211

\bibitem[\protect\citeauthoryear{{Pinzke} \& {Pfrommer}}{{Pinzke} \&
  {Pfrommer}}{2010}]{2010MNRAS.409..449P}
{Pinzke} A.,  {Pfrommer} C.,  2010, \mnras, 409, 449

\bibitem[\protect\citeauthoryear{{Planck Collaboration}}{{Planck
  Collaboration}}{2012}]{2012arXiv1208.3611P}
{Planck Collaboration} 2012, ArXiv e-prints

\bibitem[\protect\citeauthoryear{{Price}}{{Price}}{2007}]{2007arXiv0709.2772P}
{Price} D.~J.,  2007, ArXiv e-prints, 709

\bibitem[\protect\citeauthoryear{{Ricker} \& {Sarazin}}{{Ricker} \&
  {Sarazin}}{2001}]{2001ApJ...561..621R}
{Ricker} P.~M.,  {Sarazin} C.~L.,  2001, \apj, 561, 621

\bibitem[\protect\citeauthoryear{{Rybicki} \& {Lightman}}{{Rybicki} \&
  {Lightman}}{1986}]{1986rpa..book.....R}
{Rybicki} G.~B.,  {Lightman} A.~P.,  1986, {Radiative Processes in
  Astrophysics}.
Radiative Processes in Astrophysics, by George B.~Rybicki, Alan P.~Lightman,
  pp.~400.~ISBN 0-471-82759-2.~Wiley-VCH , June 1986.

\bibitem[\protect\citeauthoryear{{Ryle} \& {Windram}}{{Ryle} \&
  {Windram}}{1968}]{1968MNRAS.138....1R}
{Ryle} M.,  {Windram} M.~D.,  1968, \mnras, 138, 1

\bibitem[\protect\citeauthoryear{{Ryu}, {Kang}, {Cho} \& {Das}}{{Ryu}
  et~al.}{2008}]{2008Sci...320..909R}
{Ryu} D.,  {Kang} H.,  {Cho} J.,    {Das} S.,  2008, Science, 320, 909

\bibitem[\protect\citeauthoryear{{Sarazin}}{{Sarazin}}{1999}]{1999astro.ph.114%
39S}
{Sarazin} C.~L.,  1999, ArXiv Astrophysics e-prints

\bibitem[\protect\citeauthoryear{{Schlickeiser}}{{Schlickeiser}}{2002}]{2002cr%
a..book.....S}
{Schlickeiser} R.,  2002, {Cosmic Ray Astrophysics}.
Cosmic ray astrophysics / Reinhard Schlickeiser, Astronomy and Astrophysics
  Library; Physics and Astronomy Online Library.~Berlin: Springer.~ISBN
  3-540-66465-3, 2002, XV + 519 pp.

\bibitem[\protect\citeauthoryear{{Schlickeiser}, {Sievers} \&
  {Thiemann}}{{Schlickeiser} et~al.}{1987}]{1987A&A...182...21S}
{Schlickeiser} R.,  {Sievers} A.,    {Thiemann} H.,  1987, \aap, 182, 21

\bibitem[\protect\citeauthoryear{{Schuecker}, {Finoguenov}, {Miniati},
  {B{\"o}hringer} \& {Briel}}{{Schuecker} et~al.}{2004}]{2004A&A...426..387S}
{Schuecker} P.,  {Finoguenov} A.,  {Miniati} F.,  {B{\"o}hringer} H.,
  {Briel} U.~G.,  2004, \aap, 426, 387

\bibitem[\protect\citeauthoryear{{Spangler}}{{Spangler}}{1986}]{1986PhFl...29.%
2535S}
{Spangler} S.~R.,  1986, Physics of Fluids, 29, 2535

\bibitem[\protect\citeauthoryear{{Spitkovsky}}{{Spitkovsky}}{2008}]{2008ApJ...%
682L...5S}
{Spitkovsky} A.,  2008, \apjl, 682, L5

\bibitem[\protect\citeauthoryear{{Spitzer}}{{Spitzer}}{1956}]{1956pfig.book...%
..S}
{Spitzer} L.,  1956, {Physics of Fully Ionized Gases}

\bibitem[\protect\citeauthoryear{{Springel}}{{Springel}}{2005}]{2005MNRAS.364.%
1105S}
{Springel} V.,  2005, \mnras, 364, 1105

\bibitem[\protect\citeauthoryear{{Springel}, {Di Matteo} \&
  {Hernquist}}{{Springel} et~al.}{2005}]{2005ApJ...620L..79S}
{Springel} V.,  {Di Matteo} T.,    {Hernquist} L.,  2005, \apjl, 620, L79

\bibitem[\protect\citeauthoryear{{Subramanian}, {Shukurov} \&
  {Haugen}}{{Subramanian} et~al.}{2006}]{2006MNRAS.366.1437S}
{Subramanian} K.,  {Shukurov} A.,    {Haugen} N.~E.~L.,  2006, \mnras, 366,
  1437

\bibitem[\protect\citeauthoryear{{Sunyaev} \& {Zeldovich}}{{Sunyaev} \&
  {Zeldovich}}{1980}]{1980ARA&A..18..537S}
{Sunyaev} R.~A.,  {Zeldovich} I.~B.,  1980, \araa, 18, 537

\bibitem[\protect\citeauthoryear{{Thierbach}, {Klein} \&
  {Wielebinski}}{{Thierbach} et~al.}{2003}]{2003A&A...397...53T}
{Thierbach} M.,  {Klein} U.,    {Wielebinski} R.,  2003, \aap, 397, 53

\bibitem[\protect\citeauthoryear{{Vazza}, {Br{\"u}ggen}, {Gheller} \&
  {Brunetti}}{{Vazza} et~al.}{2012}]{2012MNRAS.421.3375V}
{Vazza} F.,  {Br{\"u}ggen} M.,  {Gheller} C.,    {Brunetti} G.,  2012, \mnras,
  421, 3375

\bibitem[\protect\citeauthoryear{{Vazza}, {Brunetti} \& {Gheller}}{{Vazza}
  et~al.}{2008}]{2008arXiv0808.0609V}
{Vazza} F.,  {Brunetti} G.,    {Gheller} C.,  2008, ArXiv e-prints, 808

\bibitem[\protect\citeauthoryear{{Vazza}, {Brunetti}, {Gheller}, {Brunino} \&
  {Br{\"u}ggen}}{{Vazza} et~al.}{2011}]{2011A&A...529A..17V}
{Vazza} F.,  {Brunetti} G.,  {Gheller} C.,  {Brunino} R.,    {Br{\"u}ggen} M.,
  2011, \aap, 529, A17+

\bibitem[\protect\citeauthoryear{{Vazza}, {Brunetti}, {Kritsuk}, {Wagner},
  {Gheller} \& {Norman}}{{Vazza} et~al.}{2009}]{2009A&A...504...33V}
{Vazza} F.,  {Brunetti} G.,  {Kritsuk} A.,  {Wagner} R.,  {Gheller} C.,
  {Norman} M.,  2009, \aap, 504, 33

\bibitem[\protect\citeauthoryear{{Venturi}}{{Venturi}}{2011}]{2011MmSAI..82..4%
99V}
{Venturi} T.,  2011, \memsai, 82, 499

\bibitem[\protect\citeauthoryear{{Venturi}, {Giacintucci}, {Brunetti},
  {Cassano}, {Bardelli}, {Dallacasa} \& {Setti}}{{Venturi}
  et~al.}{2007}]{2007A&A...463..937V}
{Venturi} T.,  {Giacintucci} S.,  {Brunetti} G.,  {Cassano} R.,  {Bardelli} S.,
   {Dallacasa} D.,    {Setti} G.,  2007, \aap, 463, 937

\bibitem[\protect\citeauthoryear{{Venturi}, {Giovannini} \&
  {Feretti}}{{Venturi} et~al.}{1990}]{1990AJ.....99.1381V}
{Venturi} T.,  {Giovannini} G.,    {Feretti} L.,  1990, \aj, 99, 1381

\bibitem[\protect\citeauthoryear{{Veritas Collaboration}, {Pfrommer} \&
  {Pinzke}}{{Veritas Collaboration} et~al.}{2012}]{2012ApJ...757..123A}
{Veritas Collaboration} {Pfrommer} C.,    {Pinzke} A.,  2012, \apj, 757, 123

\bibitem[\protect\citeauthoryear{{V{\"o}lk}}{{V{\"o}lk}}{1973}]{1973Ap&SS..25.%
.471V}
{V{\"o}lk} H.~J.,  1973, \apss, 25, 471

\bibitem[\protect\citeauthoryear{{V{\"o}lk} \& {Atoyan}}{{V{\"o}lk} \&
  {Atoyan}}{1999}]{1999APh....11...73V}
{V{\"o}lk} H.~J.,  {Atoyan} A.~M.,  1999, Astroparticle Physics, 11, 73

\bibitem[\protect\citeauthoryear{{Weinstock}}{{Weinstock}}{1969}]{1969PhFl...1%
2.1045W}
{Weinstock} J.,  1969, Physics of Fluids, 12, 1045

\bibitem[\protect\citeauthoryear{{Weinstock}}{{Weinstock}}{1970}]{1970PhFl...1%
3.2308W}
{Weinstock} J.,  1970, Physics of Fluids, 13, 2308

\bibitem[\protect\citeauthoryear{{Wentzel}}{{Wentzel}}{1974}]{1974ARA&A..12...%
71W}
{Wentzel} D.~G.,  1974, \araa, 12, 71

\bibitem[\protect\citeauthoryear{{Widrow}}{{Widrow}}{2002}]{2002RvMP...74..775%
W}
{Widrow} L.~M.,  2002, Reviews of Modern Physics, 74, 775

\bibitem[\protect\citeauthoryear{{Willson}}{{Willson}}{1970}]{1970MNRAS.151...%
.1W}
{Willson} M.~A.~G.,  1970, \mnras, 151, 1

\bibitem[\protect\citeauthoryear{{Yan} \& {Lazarian}}{{Yan} \&
  {Lazarian}}{2008}]{2008ApJ...673..942Y}
{Yan} H.,  {Lazarian} A.,  2008, \apj, 673, 942

\bibitem[\protect\citeauthoryear{{Zandanel}, {Pfrommer} \& {Prada}}{{Zandanel}
  et~al.}{2012}]{2012arXiv1207.6410Z}
{Zandanel} F.,  {Pfrommer} C.,    {Prada} F.,  2012, ArXiv e-prints

\end{thebibliography}

\end{document}